\documentclass[twocolumn]{aastex631}

\usepackage{amsmath,amssymb,amsfonts}
\usepackage{multirow} 
\usepackage{txfonts}\usepackage{soul}

\submitjournal{ApJ}

\begin{document}

\title{A Physically Driven Parameterisation of Multidimensional Atmospheres: \\
Application to the JWST Phase Curve of WASP-121b}

\correspondingauthor{Guo Chen}
\email{guochen@pmo.ac.cn}

\author[0009-0001-5682-5015]{Yuanheng Yang}
\affiliation{CAS Key Laboratory of Planetary Sciences, Purple Mountain Observatory, Chinese Academy of Sciences, Nanjing 210023, China}
\affiliation{School of Astronomy and Space Science, University of Science and Technology of China, Hefei 230026, China}

\author[0000-0003-0740-5433]{Guo Chen}
\affiliation{CAS Key Laboratory of Planetary Sciences, Purple Mountain Observatory, Chinese Academy of Sciences, Nanjing 210023, China}

\author[0000-0003-2278-6932]{Xianyu Tan}
\affiliation{Tsung-Dao Lee Institute \& School of Physics and Astronomy, Shanghai Jiao Tong University, Shanghai 201210, China}

\author[0000-0002-9258-5311]{Thaddeus D. Komacek}
\affiliation{Department of Physics (Atmospheric, Oceanic and Planetary Physics), University of Oxford, Oxford OX1 3PU, UK}

\author{Xi Zhang}
\affiliation{Department of Earth and Planetary Sciences, University of California, Santa Cruz, CA 95064, USA}

\author[0000-0001-5442-1300]{Thomas M. Evans-Soma}
\affiliation{School of Information \& Physical Sciences, University of Newcastle, Callaghan, NSW, Australia}
\affiliation{Max-Planck-Institut f\"ur Astronomie, Heidelberg, Germany}

\author[0000-0001-9585-9034]{Fei Yan}
\affiliation{Department of Astronomy, University of Science and Technology of China, Hefei 230026, China}

\author[0000-0003-1381-5527]{Chengzi Jiang}
\affiliation{Instituto de Astrof\'isica de Canarias, V\'ia L\'actea s/n, 38205 La Laguna, Tenerife, Spain}
\affiliation{Departamento de Astrof\'isica, Universidad de La Laguna, C/ Padre Herrera, 38206 La Laguna, Tenerife, Spain}

\author[0000-0002-8958-0683]{Fei Dai}
\affiliation{Institute for Astronomy, University of Hawai‘i, 2680 Woodlawn Drive, Honolulu, HI 96822, USA}



\begin{abstract}

Understanding the multidimensional structure of strongly irradiated exoplanets is essential for interpreting their atmospheric dynamics, chemistry and energy transport, yet current analyses remain limited by the difficulty of extracting reliable phase-resolved spectra and by the lack of physically interpretable parameterisations for retrievals. We present a framework combining a data-driven eclipse-normalisation method for phase-resolved emission spectra with an analytical three-dimensional temperature parameterisation, derived from radiative, advective and diffusive energy balance and controlled by a small set of characteristic timescales. Applied to JWST/NIRSpec G395H observations of WASP-121b, the method yields spectra consistent with conventional phase-curve fitting, while the parameterisation reproduces the large-scale thermal structures predicted by general circulation models. The preferred retrieval reveals a pronounced day--night contrast, a dayside thermal inversion extending to both limbs, an inversion over part of the nightside, and limb temperatures differing by several hundred kelvin. Dynamical transport strengthens with pressure, and the hotspot offset increases from $\sim4^\circ$ to $\sim9^\circ$ across the pressures probed by G395H. The confined dayside hot region and the small, pressure-dependent offsets lie closer to the $\sim$3~G GCM than to its non-magnetic counterpart, although Rayleigh drag cannot be excluded. The spectra also favour distinct dayside and nightside chemical states, with more nightside CH$_4$ than the cooler temperatures alone can explain, pointing to disequilibrium chemistry. The retrieved thermal structure further implies an inhomogeneous cloud distribution, with condensation favoured on the nightside and cooler morning limb. The framework provides a computationally efficient, physically interpretable path from spectroscopic phase curves to multidimensional atmospheric structure.

\end{abstract}


\keywords{\href{http://astrothesaurus.org/uat/487}{Exoplanet atmospheres (487)}; 
\href{http://astrothesaurus.org/uat/2021}{Exoplanet atmospheric composition (2021)}; 
\href{http://astrothesaurus.org/uat/753}{Hot Jupiters (753)}; 
\href{http://astrothesaurus.org/uat/498}{Exoplanets (498)};
\href{http://astrothesaurus.org/uat/509}{Extrasolar gaseous giant planets (509)}
}


\section{Introduction}

\subsection{Towards Multidimensional Characterisation of Exoplanet Atmospheres}

Characterising the multidimensional atmospheric structure has become a central objective in exoplanet science, particularly in the era of James Webb Space Telescope (JWST) and Ariel \citep{Tinetti+etal+2018}. Over the past two decades, rapid progress in both observations and modelling \citep[e.g.,][]{Madhusudhan+etal+2019} has enabled studies that extend beyond one-dimensional (1D) or disk-averaged descriptions towards genuinely three-dimensional (3D) atmospheric characterisation \citep[e.g.,][]{Parmentier+etal+2018a,Pluriel+etal+2023}. A key goal is to constrain large-scale circulation, spatially varying chemistry, and cloud distributions, together with the physical mechanisms that govern them.

At the precision enabled by JWST, hot Jupiters have become critical laboratories for developing and validating methodologies aimed at inferring multidimensional atmospheric structures. Historically, 1D analyses have characterised how strongly irradiated atmospheres absorb and reemit stellar energy, constraining their vertical thermal structure through parameterised temperature-pressure profiles and simplified prescriptions for opacities of chemical composition \citep[e.g.,][]{Hansen+etal+2008,Madhusudhan+etal+2009,Guillot+etal+2010,Line+etal+2013,Pelletier+etal+2021}. These formulations provided computationally tractable and observationally commensurate approaches during the pre-JWST era. However, characterising the multidimensional atmospheric structure requires addressing the full set of radiative, dynamical, and thermochemical processes that regulate the global redistribution of energy and the transport of momentum and mass under extreme irradiation. Transitioning from 1D models to frameworks that capture the inherently 3D nature of hot-Jupiter circulation constitutes a key challenge for next-generation atmospheric models and retrievals. Moreover, several studies have investigated the systematic biases that arise from adopting 1D assumptions in atmospheric modelling and retrieval instead of considering the inherently 3D nature of atmospheric structures \citep[e.g.,][]{Feng+etal+2016, Pluriel+etal+2020, MacDonald+etal+2020, Arora+etal+2025}

Although the inherently multidimensional nature of planetary atmospheres is widely recognised, relatively few studies have attempted to retrieve the intrinsic atmospheric processes of tidally locked giant planets or to assess their fundamental dynamical properties in a manner directly comparable to observations. In recent years, substantial progress has been made in the development of general circulation models (GCMs) that characterise the 3D structure of exoplanet atmospheres \citep[e.g.,][]{Showman+etal+2009,2024MNRAS.528.1016T,Heng+etal+2011,Cho+etal+2021,2019ApJ...886...26T,Komacek+etal+2022,Roth+etal+2024}. Nevertheless, establishing a direct correspondence between the physically controlled parameters or core dynamical features of GCMs and the observables remains difficult. This challenge arises largely from the historical absence of physically motivated, user-accessible, parameter-controllable frameworks capable of describing multidimensional atmospheric structure within a retrieval context, particularly temperature fields. 

Several exploratory approaches have begun incorporating multidimensional atmospheric structures into retrieval frameworks. These efforts can broadly be divided into two categories. The first category includes freely parameterised structural models that introduce additional degrees of freedom to describe multidimensional temperature fields \citep{Changeat+etal+2019,MacDonald+etal+2020,MacDonald+etal+2022,Feng+etal+2016,Feng+etal+2020}. \citet{Irwin+etal+2020} incorporated a cosine-based latitudinal dependence when computing disk-integrated spectra, and \citet{Yang+etal+2023} examined several parametric schemes for two-dimensional temperature distributions. \citet{Nixon+etal+2022} introduced AURA-3D, which combines a rapid 3D transmission spectrum with a parametric 3D pressure-temperature structure designed to reproduce azimuthally averaged temperature structures. \citet{Zingales+etal+2022} developed TauREx 2D, which uses a two-dimensional limb parameterisation to account for day-night thermal differences in transmission spectra. The second category consists of physically driven parameterisations, which derive multidimensional structural forms and associated control parameters from GCMs or from simplified conservation equations. For example, \citet{Chubb+etal+2022} obtained two-dimensional temperature distributions by iteratively solving radiative-diffusive equilibrium on a latitude-longitude plane, and \citet{Lewis+etal+2022} analysed distinct components of the thermal structure. While these approaches are physically grounded, their computational complexity and lack of analytic closed-form solutions hinder their use in retrieval frameworks. \citet{Dobbs-Dixon+etal+2022} proposed an analytic parameterisation inspired by GCM temperature structures that preserves key large-scale physical behaviour, offering a promising alternative.

For characterising atmospheric multidimensionality, particularly in hot Jupiters, a physically driven family of multidimensional temperature field parameterisations is essential. These parameterisations must be capable of representing large-scale structures with well-defined control parameters and be readily incorporated into retrieval frameworks for diverse observational data (e.g., spectroscopic phase curves, eclipse mapping, transmission spectroscopy). \citet{2002A&A...385..166S} introduced characteristic timescales for tidally locked atmospheres, such as radiative and advective timescales, which provide a physically motivated simplification of the circulation. Constraining these timescales enables quantitative assessment of large-scale circulation properties. This includes determining whether a given layer is dominated by radiative heating and cooling or by advection, evaluating the efficiency of day-night energy redistribution, and analysing the distribution of energy across latitude, longitude, and pressure. Given current limitations in data quality and observational techniques, it is practical to describe the multidimensional structure of atmospheres by parameterising their large-scale temperature structure using a small set of governing timescales. However, finer-scale characterisation will require improved data and reduced observational degeneracies, such as the latitudinal degeneracies inherent in phase-resolved spectra.


In this work, we develop an analytical temperature parameterisation theory derived from the radiative-advective-diffusive equilibrium equation, expressed in terms of a small set of physically motivated control parameters (i.e., various characteristic timescales). The resulting formulation captures the large-scale, multidimensional temperature structure of the atmosphere and can be readily incorporated into retrievals, as demonstrated here using WASP-121b spectroscopic phase-curve observations by JWST/NIRSpec G395H. This physically driven parameterisation provides a transparent and interpretable representation of the atmospheric structure, offering new quantitative constraints and insights into the multidimensional, large-scale thermal structures of close-in tidally locked planetary atmospheres.

\subsection{Data-driven Extraction of Spectroscopic Phase Curves}

Phase-resolved emission spectroscopy constitutes one of the most informative observational techniques for characterising the multidimensional atmospheric structure of strongly irradiated exoplanets. Such observations provide constraints on their global energy balance, large-scale circulation, and chemical inhomogeneities \citep[e.g.,][]{Knutson+etal+2007,Cowan+etal+2008,Stevenson+etal+2014,Challener+etal+2025,Mikal-Evans+etal+2022,Mikal-Evans+etal+2023}. For tidally locked planets, the orbital phase directly corresponds to the rotational phase. This allows us to infer longitudinally resolved information from the observed flux modulation. In the infrared, the amplitude of the phase variation constrains the day-night temperature contrast. Meanwhile, the phase offset of the flux maximum reveals the longitudinal displacement of the atmospheric hotspot. This displacement is typically associated with eastward equatorial jets that transport energy from the intensely irradiated dayside to the cooler nightside. Traditionally, extracting phase-resolved emission spectra from time-series observations requires fitting spectroscopic phase curves with phase curve models. Although this procedure is widely adopted and physically well motivated, it typically requires explicit assumptions about the form of the phase variation and joint modelling of instrumental systematics, which can increase the complexity of the modelling.

Here, we employ a data-driven, end-to-end methodology to derive phase-resolved emission spectra directly from JWST time-series observations, without explicitly fitting phase-curve models. By directly reconstructing wavelength-dependent phase variations from the time series, this ``eclipse normalisation'' approach provides a transparent and computationally efficient way to extract spectroscopic phase curves while retaining the accuracy required for atmospheric characterisation. The approach is useful for datasets in which instrumental systematics are minimal. We note that \citet{Lustig-Yaeger+etal+2025} independently developed a similar method based on the same concept. A closely related direct-extraction strategy was applied to JWST/NIRISS observations of WASP-18b by \citet{Ouyang+etal+2026}. Furthermore, \citet{Changeat+etal+2024} advocated for the adoption of data-driven processing and retrieval approaches in the JWST and Ariel, underscoring the growing need for flexible processing methodologies.

\subsection{Outline}

This paper is organised as follows. Section~\ref{METHOD_EN} introduces the eclipse normalisation method to extract phase-resolved emission spectra. Section~\ref{sec:model} presents the new analytical temperature parameterisation framework, assesses its performance, and outlines the associated forward-model calculations. Section~\ref{RETR} presents the retrieval framework and results for spectroscopic phase-curve observations of WASP-121b with JWST/NIRSpec G395H. The discussion and conclusions are presented in Section~\ref{DISC} and Section~\ref{CONCL}, respectively.

\section{Data-Driven Phase Curve Extraction}\label{METHOD_EN}
We utilise the full-orbit phase curve observations of WASP-121b obtained with the JWST/NIRSpec G395H grating (Program GO-1729; PI: Mikal-Evans, Co-PI: Kataria). The observation window spans two consecutive secondary eclipses, capturing the planet's complete thermal emission cycle. While previous studies \citep{Mikal-Evans+etal+2023, Evans-Soma+etal+2025} have analysed this dataset using standard light-curve fitting to constrain dayside and nightside compositions, we present here a data-driven methodology. This approach allows for the extraction of phase-resolved spectra directly from the time series, without explicitly fitting parametric phase-curve models, and serves as a complementary quick-look analysis alongside traditional light-curve fitting.

\subsection{Treating Phase Curves as High-Resolution Spectra}
\label{sec:data_ana}
Inspired by high-resolution spectroscopy, we develop a pipeline that constructs the phase-resolved spectral matrix directly from the time series spectra following reduction by the FIREFLy suite, obtained from \mbox{\citet{Evans-Soma+etal+2025}}. This method eliminates the need for traditional light-curve fitting. The core concept involves isolating the planetary phase signal (expressed as 1+$F_\mathrm{p}/F_\mathrm{s}$) by dividing each spectrum in the time series by a master stellar spectrum. This master spectrum is derived by averaging all spectra obtained during the full secondary eclipse phases (i.e., excluding ingress and egress). We term this method ``eclipse normalisation,'' and its implementation proceeds in the following stages.

The JWST/NIRSpec G395H observations utilise two detectors, NRS1 and NRS2, covering the wavelength ranges of 2.70-3.72 $\micron$ and 3.82-5.15 $\micron$, respectively. We first concatenate the spectral data from both detectors to form a continuous spectral matrix. To ensure data quality, we perform a two-step cleaning process. First, outlier integrations are identified by constructing a white-light curve (summed over the full wavelength range). We apply an 11-point median filter to define the baseline and discard any integrations deviating by more than $5\sigma$, where $\sigma$ is estimated via the median absolute deviation. Second, we address pixel-level artifacts. While the FIREFLy suite \citep{Rustamkulov+etal+2022,Rustamkulov+etal+2023} provides initial bad pixel flagging, we refine this by removing wavelength columns that are entirely flagged (along with their neighbours). Isolated bad pixels are corrected via linear interpolation along the time axis. Finally, a robust iterative $\sigma$-clipping procedure is applied to the entire matrix, alternating between column-wise clipping and interpolation until convergence, effectively suppressing any residual outliers.

\begin{figure}
\includegraphics[trim=0.75cm 0.35cm 0.2cm 0.2cm, width=1\linewidth]{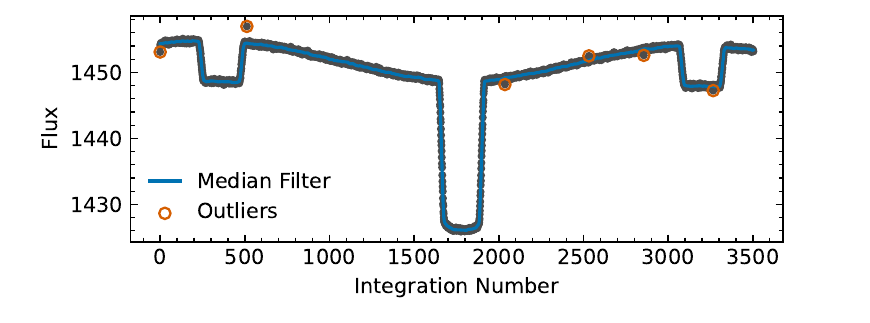}
\includegraphics[trim=0.18cm 0.35cm 0.62cm 0.0cm,width=1\linewidth]{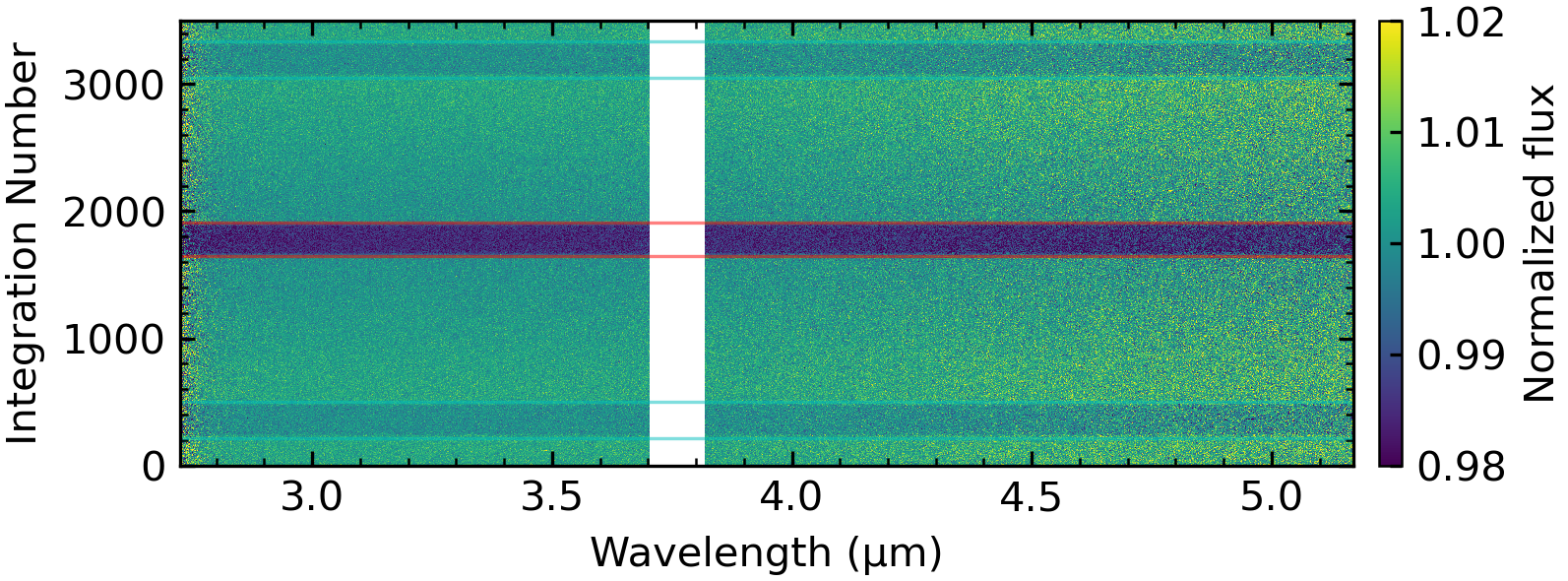}
\caption{Data cleaning and normalisation process. Top: Identification and removal of outlier integrations based on the white-light curve. Bottom: The resulting cleaned and concatenated phase-resolved spectral matrix, showing the normalised flux ratio at the native instrumental resolution.
}
\label{Fig1}
\end{figure}

Following preprocessing, we isolate the integrations obtained during the full secondary eclipse (i.e., excluding ingress and egress) to derive the pure stellar spectrum, $F_\mathrm{s}(\lambda)$. The master stellar template is defined as the median of these fully-in-eclipse spectra, with its uncertainty estimated as the standard error of the median ($\sigma_{\rm ecl}(\lambda)/\sqrt{N_{\rm ecl}}$). Each cleaned integration in the time series is then divided by this master template to generate the normalised residual matrix, $R(\varphi, \lambda) = F(\varphi, \lambda)/F_\mathrm{s}(\lambda)$, where $\varphi$ denotes the orbital phase angle. The associated uncertainties are propagated in quadrature, accounting for both the photon noise of individual integration and the uncertainty of the master template. Both the residual spectral matrix and the corresponding uncertainty estimates are preserved at the native instrumental resolution for subsequent analysis, as illustrated in Figure~\ref{Fig1}. 

\begin{figure*}
\centering
\includegraphics[width=\textwidth, trim=10 10 10 10]{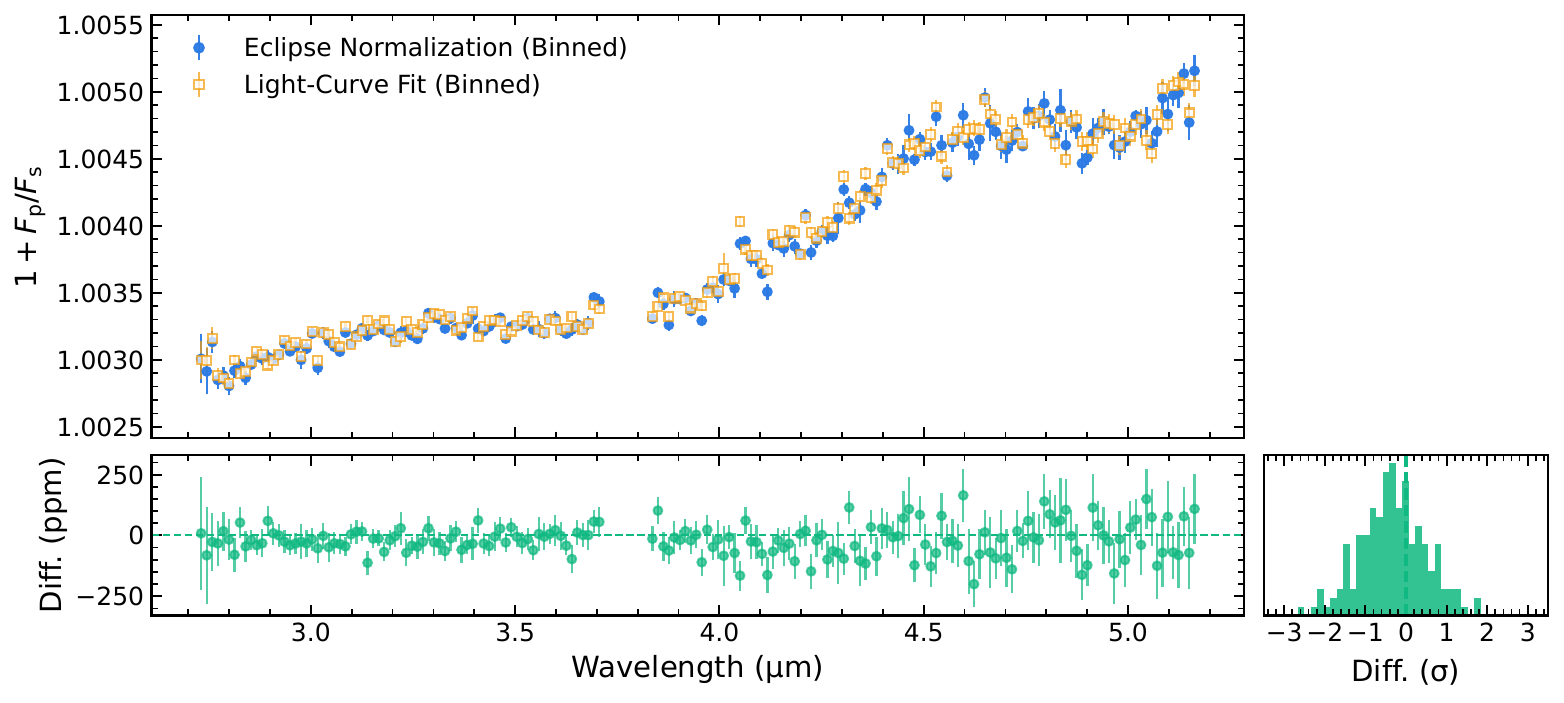}
\caption{Comparison of dayside thermal emission spectra obtained via eclipse normalisation and standard light-curve fitting. Top: Overplotted spectra from both methods. Bottom: Residuals (difference) between the two spectra and the histogram of their distribution. 
}
\label{Fig2}
\end{figure*}

\begin{figure*}
\centering
\includegraphics[width=\textwidth, trim=10 10 10 10]{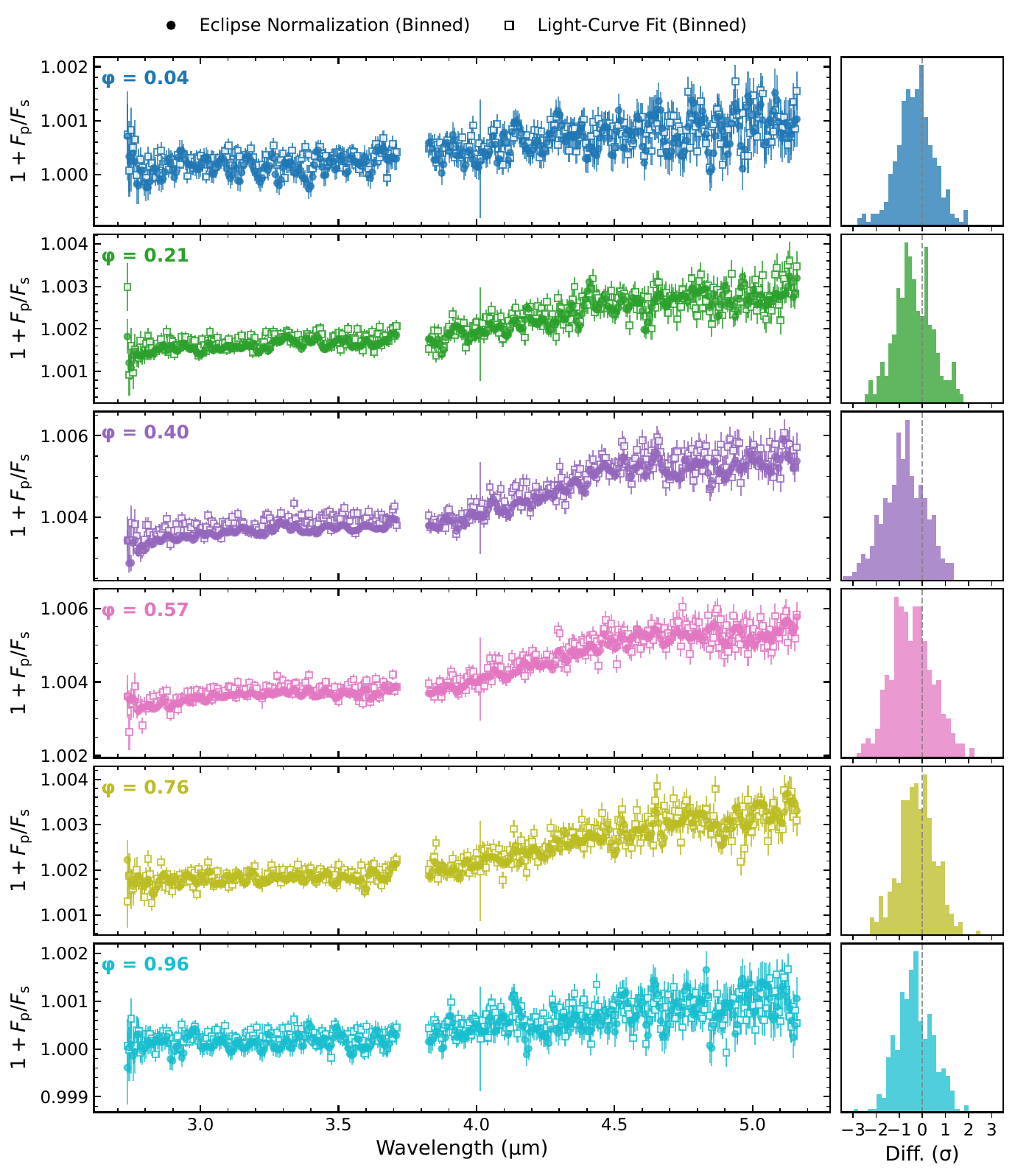}
\caption{Comparison of phase-binned thermal emission spectra obtained via eclipse normalisation and standard light-curve fitting. 
}
\label{Fig3}
\end{figure*}

\subsection{Accounting for Instrumental Systematics}
\label{sec:model_sys}

The normalised residual matrix, $R(\varphi, \lambda)$, contains both the astrophysical signal encoding the planetary thermal emission and a residual term representing time- and wavelength-dependent instrumental systematics. Prior to JWST, complex instrumental systematics and limited photometric precision rendered the eclipse normalisation approach impractical. Consequently, for low- and medium-resolution spectroscopy, recovering planetary spectra typically necessitated parametric light-curve fitting, which jointly models astrophysical signals and instrumental systematics. However, the unprecedented stability and precision of JWST have made these systematics far more tractable. This shift has revitalised eclipse normalisation as a first-principles methodology, allowing us to preserve the original spectral information with minimal reliance on complex reduction steps. 

Within the eclipse normalisation framework, we identify two primary strategies for addressing instrumental systematics. The first approach involves modelling systematics jointly with the planetary atmosphere during the retrieval process. This requires a comprehensive understanding of the physical origins and parametric forms of the systematic noise. The second approach involves correcting for systematics prior to the normalisation step. For JWST phase-curve observations covering two secondary eclipses, the eclipses themselves can serve as a reference baseline to characterise and remove systematic trends. However, this method assumes that the host star is photometrically stable with negligible variability. In cases involving late-type stars, where stellar activity is significant, such variability must be explicitly modelled and incorporated into the retrieval framework alongside the planetary atmosphere, necessitating the first approach. 

In this study, we use the two secondary eclipses captured in the WASP-121b phase curve to characterise and remove instrumental trends. Specifically, we use the in-eclipse data as a baseline to fit a linear time-dependent function for each wavelength channel, a model shown to effectively describe JWST/NIRSpec systematics \citep{Mikal-Evans+etal+2023,Evans-Soma+etal+2025}. The uncertainties associated with these fitted trends are propagated into the residual spectra. Regarding stellar activity, WASP-121 is a slowly rotating F-type star with minimal activity \citep{Delrez+etal+2016}, implying negligible flux variability over the observation window. Consequently, the stellar spectrum can be treated as constant within measurement uncertainties. This allows us to proceed with atmospheric retrieval without the need for additional parameters to model complex instrumental systematics or stellar activity.

\subsection{Validating against Standard Light-Curve Fitting}

Figure~\ref{Fig2} compares the dayside thermal emission spectra derived using two different approaches.
The eclipse-normalised spectrum is obtained using our method, whereas the spectrum derived from standard light-curve fitting is reported by \mbox{\citet{Evans-Soma+etal+2025}}. Specifically, the eclipse-normalised dayside spectrum is computed by averaging the phase-resolved spectra over the phase interval corresponding to the ``dayside,'' defined to be identical to the phase range used in the standard light-curve fitting. In contrast to full-orbit analyses that infer the stellar flux, planetary phase variation, and instrumental response jointly \citep[e.g.,][]{Cowan+etal+2012}, eclipse normalisation anchors the stellar spectrum directly to the two full secondary eclipses. Figure~\ref{Fig3} further presents the extracted thermal emission spectra binned at selected orbital phases. The two methods yield mutually consistent results for this dataset, indicating that eclipse normalisation captures the dominant phase-dependent signal in the JWST/NIRSpec observations of WASP-121b. Compared to traditional light-curve fitting, the eclipse-normalisation approach involves substantially fewer model assumptions during the extraction stage and is computationally lightweight. Quantitatively, the residuals between the two methods lie entirely within the 3$\sigma$ confidence interval, consistent with statistical agreement between the approaches.

\section{Atmospheric Modelling Framework}\label{sec:model}

\subsection{3D Temperature Parameterisation via Advection-Diffusion-Radiation Balance}
\label{sec:para_tp}

The thermodynamic energy equation commonly used in GCMs based on primitive equations can be written as
\begin{equation}\label{ther_eq1}
\frac{dT}{dt} = \frac{q}{c_p} + \frac{\omega}{\rho c_p} + \mathcal{D}_T, 
\end{equation}
where $\omega$ is the vertical velocity in pressure coordinates, $d/dt \equiv \partial/\partial t + \mathbf{v} \cdot \nabla + \omega \partial/\partial p$ is the material derivative, $q$ is the specific heating rate, and $\mathcal{D}_T$ is the additional source or sink terms. $T$, $\rho$, and $c_p$ denote the temperature, density, and specific heat at constant pressure, respectively. 

Inspired by \citet{Zhang+etal+2017} and \citet{2016ApJ...821...16K}, we construct a kinematic model. Assuming steady-state conditions and neglecting $\mathcal{D}_T$, Equation~\ref{ther_eq1} can be rewritten as: 
\begin{equation}\label{ther_eq2}
\mathbf{v} \cdot \nabla T + \omega \frac{\partial T}{\partial p} - \frac{\omega}{\rho c_p} = \frac{q}{c_p}.
\end{equation}
To derive an analytical solution in the longitudinal direction, we must simplify the model. However, directly neglecting vertical entropy advection (i.e., the second and third terms on the left-hand side of Equation~\ref{ther_eq2}) is arbitrary and physically unjustifiable for hot Jupiters \citep{2016ApJ...821...16K}. To preserve its role while maintaining analytical tractability, we instead parameterise the net dynamical transport represented by the left-hand side of Equation~\ref{ther_eq2} using an effective one-dimensional longitudinal transport operator, introducing a horizontal “advection” term and a “diffusion” term. In this kinematic description, the energy transport is represented by a zonal advection term together with an effective longitudinal diffusion term, serving as an approximation to the full dynamical processes represented in Equation~\ref{ther_eq2}.

The thermodynamic equation is thus reduced to a 1D form in the zonal direction (longitude $\theta$). 
We adopt the Newtonian cooling approximation for the radiative scheme, in which the radiative heating term is parameterised as
\begin{equation}
\frac{q}{c_p} = \frac{T_{\mathrm{eq}} - T}{\tau_{\rm rad}},
\end{equation}
where $\tau_{\rm rad}$ is the radiative relaxation timescale and $T_{\mathrm{eq}}$ is the equilibrium temperature.
The longitudinal energy transport is represented by a constant advective timescale $\tau_{\rm adv} \sim L / \bar{u}$, where $\bar{u}$ is the zonal-mean zonal wind, together with an effective diffusion timescale $\tau_{\rm diff}$. 
The resulting kinematic equation is:
\begin{equation}\label{ther_eq3}
\frac{1}{\tau_{\rm adv}} \frac{\partial T}{\partial \theta} - \frac{1}{\tau_{\rm diff}} \frac{\partial^2 T}{\partial \theta^2} \;=\; \frac{T_{\mathrm{eq}} - T}{\tau_{\rm rad}}.
\end{equation}

Here, $T_{\mathrm{eq}}$ is the equilibrium temperature, which we model as: 
\begin{equation}\label{ther_eq4}
T_{\rm eq}=
\begin{cases}
T_n+\Delta T\cos\phi\Big[(1-\alpha)+\alpha\cos\theta\Big],
& \text{dayside},\\
T_n, & \text{nightside},
\end{cases}
\end{equation}
where $\phi$ is the latitude, $T_{\rm n}$ is the temperature at the antistellar point, and $\Delta T$ is the temperature contrast between the substellar and antistellar points. The dimensionless parameter $\alpha\in[0,1]$ controls the longitudinal sharpness of the dayside equilibrium temperature profile. The limit $\alpha=1$ recovers the standard $\cos\theta$ dependence, corresponding to a maximally contrasted dayside forcing proportional to the local insolation \citep[e.g.,][]{2016ApJ...821...16K,Zhang+etal+2017,2002A&A...385..166S}. Conversely, $\alpha=0$ yields a flat-top longitude-independent equilibrium temperature profile, representing a weak-contrast forcing. 
Our formulation, combined with a simplified $\cos\phi$ latitudinal dependence, enhances the flexibility of the parameterisation while retaining a clear physical interpretation. 

Then, we can rewrite Equation~\ref{ther_eq3} as: 
\begin{equation}\label{ther_eq5}
\varepsilon\,\frac{\partial T}{\partial \theta} - \psi\,\frac{\partial^2 T}{\partial \theta^2} \;=\; T_{\mathrm{eq}} - T,
\end{equation}
where the dimensionless parameters are defined as $\varepsilon = \tau_{\rm rad}/\tau_{\rm adv}$ and $\psi = \tau_{\rm rad}/\tau_{\rm diff}$. The general solution for the 3D temperature profile over the domain $\theta\in[-\pi,\pi]$ is given below.\footnote{The source code that implements this temperature parameterisation is available at \url{https://zenodo.org/records/21534046}.} 
\begin{equation}\label{ther_eq6}
T(\theta,\phi,p)=
\begin{cases}
\begin{aligned}
T_n(p)
&+H+A\cos\theta+B\sin\theta\\
&+C_d e^{k_1\theta}+D_d e^{k_2\theta},
\end{aligned}
& -\tfrac{\pi}{2}\le\theta\le\tfrac{\pi}{2},\\[6pt]
T_n(p)+C_l e^{k_1\theta}+D_l e^{k_2\theta},
& -\pi\le\theta\le-\tfrac{\pi}{2},\\[6pt]
T_n(p)+C_r e^{k_1\theta}+D_r e^{k_2\theta},
& \tfrac{\pi}{2}\le\theta\le\pi,
\end{cases}
\end{equation}
where
\begin{equation}\label{ther_eq7}
k_{1,2}=\frac{\varepsilon\pm\sqrt{\varepsilon^2+4\psi}}{2\psi},
\qquad k_1>0,\;k_2<0.
\end{equation}
The particular coefficients are
\begin{equation}\label{ther_eq8}
\begin{aligned}
H &= \Delta T\cos\phi\,(1-\alpha),\\
A &= \Delta T\cos\phi\,\alpha
\frac{1+\psi}{(1+\psi)^2+\varepsilon^2},\\
B &= \Delta T\cos\phi\,\alpha
\frac{\varepsilon}{(1+\psi)^2+\varepsilon^2}.
\end{aligned}
\end{equation}
Let \(\mathcal{Q}_i^{\pm}=A+k_i(B\pm H)\) for \(i=1,2\).
The homogeneous coefficients are
\begin{alignat}{2}
C_d &= \frac{\mathcal{Q}_2^+e^{k_1\pi/2}+\mathcal{Q}_2^-e^{-k_1\pi/2}}
{2(k_1-k_2)\sinh(k_1\pi)},\; &
D_d &= -\frac{\mathcal{Q}_1^+e^{k_2\pi/2}+\mathcal{Q}_1^-e^{-k_2\pi/2}}
{2(k_1-k_2)\sinh(k_2\pi)},\label{ther_eq9}\\
C_r &= C_d-\frac{\mathcal{Q}_2^+}{k_1-k_2}e^{-k_1\pi/2},\; &
D_r &= D_d+\frac{\mathcal{Q}_1^+}{k_1-k_2}e^{-k_2\pi/2},\label{ther_eq10}\\
C_l &= C_d+\frac{\mathcal{Q}_2^-}{k_1-k_2}e^{k_1\pi/2},\; &
D_l &= D_d-\frac{\mathcal{Q}_1^-}{k_1-k_2}e^{k_2\pi/2}.\label{ther_eq11}
\end{alignat}

In summary, our model parameterises the 3D temperature structure $T(\theta, \phi, p)$ using five key variables: the parameters $T_{\rm n}$ and $T_{\rm d}$, which define the equilibrium temperatures at the antistellar and substellar points in Equation~\ref{ther_eq4} (with $\Delta T = T_{\rm d} - T_{\rm n}$), together with the dimensionless parameters $\alpha$, $\varepsilon$, and $\psi$. Since Equation~\ref{ther_eq5} is independent of latitude and pressure, we can freely prescribe the pressure and latitudinal dependence of these parameters without compromising analyticity. It is worth noting that, if one wishes to relax the latitudinal dependence, Equation~\ref{ther_eq4} no longer requires the commonly adopted $\cos\phi$ dependence of $T_{\rm eq}$ in latitude. To maintain a manageable number of free parameters for retrieval, we assume that all latitudes share a common set of \(T_{\mathrm{n}}\), \(T_{\mathrm{d}}\), \(\alpha\), \(\varepsilon\), and \(\psi\). In the pressure dimension, \(T_{\mathrm{n}}(p)\) and \(T_{\mathrm{d}}(p)\) can be described by arbitrary parametric profiles \citep[e.g.,][]{Guillot+etal+2010,Pelletier+etal+2021}, while \(\alpha(p)\), \(\varepsilon(p)\) and \(\psi(p)\) can be parameterised using flexible functional forms (e.g., linear or power-law).

In addition, we define the hotspot longitude, $\theta_{\rm h}$, as the location where the longitudinal temperature gradient vanishes,
\begin{equation}\label{ther_eq12}
  \left.\frac{\partial T}{\partial\theta}\right|_{\theta=\theta_{\rm h}} = 0,
  \qquad -\frac{\pi}{2}\le\theta_{\rm h}\le\frac{\pi}{2}.
\end{equation}
Substituting this into Equation~\ref{ther_eq6}, the hotspot offset is therefore the solution of the equation
\begin{equation}\label{ther_eq13}
  -A\sin\theta_{\rm h} + B\cos\theta_{\rm h}
  + k_1 C_d e^{k_1\theta_{\rm h}}
  + k_2 D_d e^{k_2\theta_{\rm h}} = 0,
\end{equation}
While no explicit analytical solution exists for $\theta_{\rm h}$, the hotspot offset is well approximated by $\theta_{\rm h} \approx \tan^{-1}\!\left(\frac{\varepsilon} {1+\psi}\right)$ in the strong-radiative regime (i.e, $\psi \ll 1$ and $\varepsilon \ll 1$). 

\subsection{Behaviour and Validation of the Temperature Parameterisation}

\begin{figure}
\centering
\includegraphics[width=0.45\textwidth, trim=10 5 10 5]{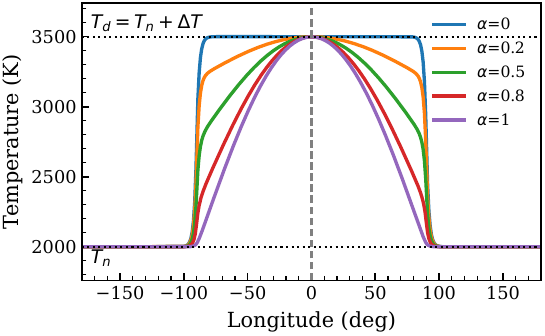}
\includegraphics[width=0.45\textwidth, trim=10 5 10 10]{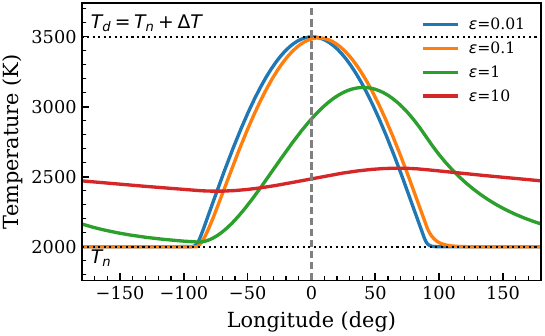}
\includegraphics[width=0.45\textwidth, trim=10 5 10 10]{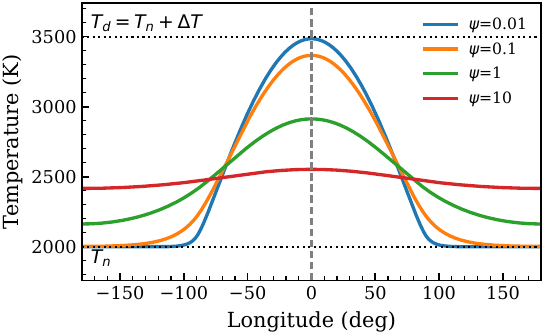}
\caption{Sensitivity of the parameterised longitudinal temperature profile to variations in the key control parameters. Top: Effect of the sharpness parameter $\alpha$. Dynamical transport is suppressed in this example by fixing  $\varepsilon=10^{-3}$ and $\psi=10^{-3}$, ensuring the temperature distribution is dominated by radiative processes. Middle: Effect of the dynamical control parameter $\varepsilon = \tau_{\rm rad}/\tau_{\rm adv}$. Other parameters are fixed ($\psi=10^{-3}$ and $\alpha=1$). Bottom: Effect of the dynamical control parameter $\psi = \tau_{\rm rad}/\tau_{\rm diff}$. Other parameters are fixed ($\varepsilon=10^{-3}$ and $\alpha=1$).
}
\label{Fig4}
\end{figure}

To describe the longitudinal variations of the equilibrium temperature $T_{\rm eq}$, our prescription (Equation~\ref{ther_eq4}) introduces a dimensionless parameter $\alpha$. As illustrated in the top panel of Figure~\ref{Fig4}, this parameter modulates the sharpness of the dayside equilibrium temperature profile, controlling the longitudinal distribution of the radiative forcing.

The efficiency and geometry of energy transport are regulated by the two dimensionless control parameters, \(\varepsilon\) and \(\psi\). The parameter \(\varepsilon\) determines the relative dominance of radiative versus advective processes, as shown in the middle panel of Figure~\ref{Fig4}. For small \(\varepsilon\), the radiative timescale is much shorter than the advective timescale, placing the atmosphere in a radiation-dominated regime. In this limit, the upper atmosphere develops a pronounced dayside hot region near the substellar point and a cold region on the nightside, resulting in a large day--night temperature contrast. Consequently, the thermal hotspot offset $\theta_{\rm h}$ is minimal because the term $\tan^{-1}\!\left(\frac{\varepsilon}{1+\psi}\right)$ remains small. Conversely, large \(\varepsilon\) values correspond to an advection-dominated regime in which efficient horizontal heat transport substantially reduces the day-night temperature contrast, driving the atmosphere towards quasi-isothermal conditions.

The bottom panel of Figure~\ref{Fig4} shows how the zonal temperature structure varies with \(\psi\). This parameter governs the efficiency of energy transport along the axis of symmetry, effectively controlling the degree of longitudinal temperature relaxation. When \(\psi\) is small (implying a long effective ``diffusion'' timescale), symmetric longitudinal transport becomes inefficient, and the temperature distribution is primarily shaped by the interplay between advection and radiative equilibrium.

\begin{figure*}
\centering
\includegraphics[width=\textwidth, trim=10 5 10 10]{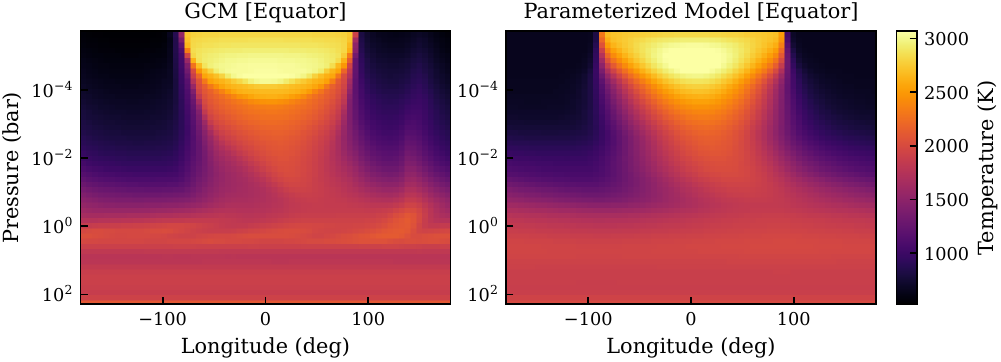}
\caption{Validation of the temperature parameterisation against GCM simulations. Left: The longitude-pressure temperature structure at the equator from the hot Jupiter GCM \citep{Roth+etal+2024}. Right: The best-fitting temperature field reconstructed using our physically driven parameterisation, characterised by the antistellar and substellar temperatures, $T_{\rm n}$ and $T_{\rm d}$, together with the dimensionless parameters $\alpha$, $\varepsilon$, and $\psi$.
}
\label{Fig5}
\end{figure*}

\begin{figure*}
\centering
\includegraphics[width=\textwidth, trim=10 5 10 10]{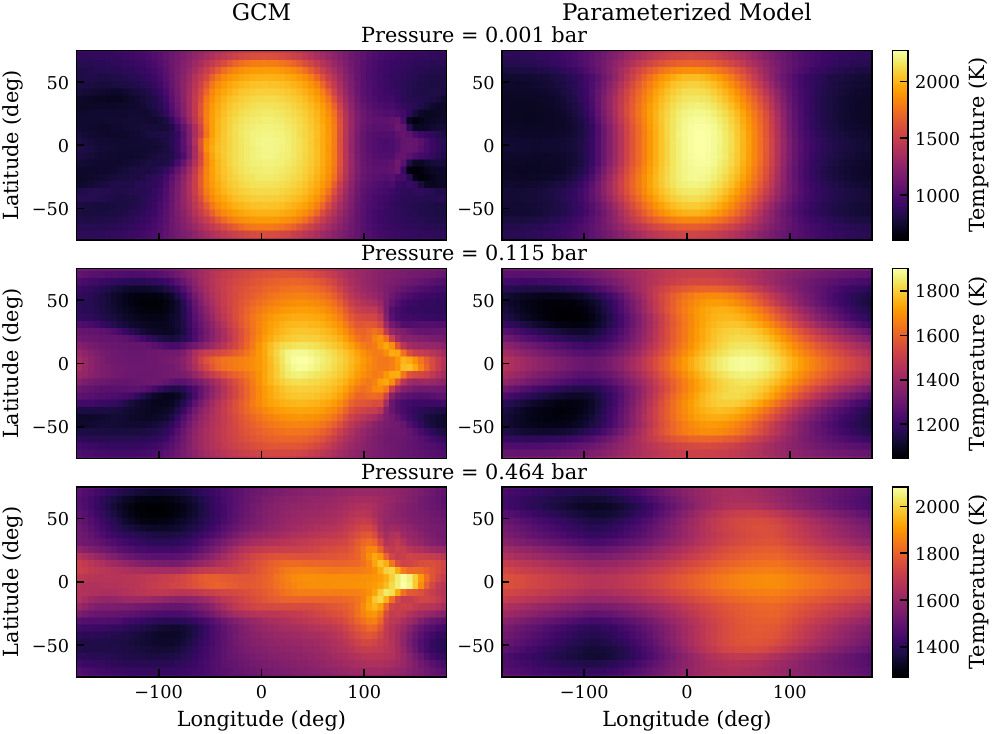}
\caption{Comparison of longitude-latitude temperature maps at selected pressure levels (0.001, 0.115, and 0.464 bar from top to bottom), between the hot Jupiter GCM (left panels) from \citet{Roth+etal+2024} and our temperature parameterisation (right panels). In this validation example, the control parameters $(\varepsilon,\psi)$ are allowed to vary with latitude to fully capture the 3D structure. 
}
\label{Fig6}
\end{figure*}

We assess the fidelity and flexibility of this framework by benchmarking it against temperature fields derived from 3D GCM simulations of representative tidally locked hot and ultra-hot Jupiters \citep[e.g.,][]{Roth+etal+2024,2024MNRAS.528.1016T}. Specifically, we reconstruct the large-scale temperature structure using our physically motivated parameter set \citep[$T_{\rm n}, T_{\rm d}, \alpha, \varepsilon, \psi$,][]{2002A&A...385..166S, 2016ApJ...821...16K, Zhang+etal+2017, Showman+etal+2020}. Figure~\ref{Fig5} presents a validation case where we fit the 3D temperature distribution of a non-grey, cloud-free, drag-free hot Jupiter GCM simulation from \citet{Roth+etal+2024}. The GCM exhibits a radiation-dominated thermal inversion on the dayside upper atmosphere, transitioning to a deep advection-dominated region with nearly isothermal day-night profiles at higher pressures. Our parameterised model successfully reproduces these key morphological features. Figure~\ref{Fig6} further confirms the agreement by comparing temperature distributions at various pressure levels.

Although our model can reproduce latitude-dependent structures by fitting different parameter sets to different latitudes (as done for the GCM benchmark), for subsequent phase-curve retrievals, we adopt a single, latitude-independent parameter set that matches the equatorial temperature structure. This effectively yields a cosine-like decline in temperature towards higher latitudes. In reality, 3D circulation exhibits both zonal flow and meridional variations, but we simplify the latitude dependence for two reasons. First, unlike eclipse mapping which offers spatial resolution \citep{Majeau+etal+2012, deWit+etal+2012, Coulombe+etal+2023, Hammond+etal+2024, Challener+etal+2025}, disk-integrated spectroscopic phase curves are highly degenerate in the latitudinal direction. Second, the equatorial region dominates the spectral projection weights \citep{Irwin+etal+2020,Parmentier+etal+2018a}, and hot-Jupiter atmospheres generally maintain higher temperatures near the equator \citep{Showman+etal+2011,Showman+etal+2020}. 

Therefore, while a full representation would require latitudinally varying $\varepsilon$ and $\psi$ (increasing computational cost and degeneracy), our simplified approach strikes an optimal balance. It avoids the excessive complexity of full GCMs while producing a multidimensional temperature field that remains physically self-consistent. This framework provides a practical pathway for constraining the 3D thermal structure of exoplanet atmospheres and interpreting large-scale temperature patterns from JWST observations. 

\subsection{Atmospheric Chemistry and Additional Assumptions}

At each atmospheric column $(\theta_i,\phi_j)$, where $i$ and $j$ index longitude and latitude grid points, we calculate molecular abundances assuming chemical equilibrium. These abundances are computed on the model pressure grid using the input overall metallicity and elemental-to-hydrogen ratios. The mean molecular weight and thermodynamic quantities are updated consistently. Equilibrium abundances are evaluated on-the-fly with \textit{FastChem} \citep{Stock+etal+2018} and mapped onto the species set used by \textit{petitRADTRANS} \citep{2019A&A...627A..67M}, ensuring consistency between chemistry, temperature, and other atmospheric properties.

In this work, the baseline (\textsc{Base}) model assumes local chemical equilibrium, spatially uniform elemental abundances, and a cloud-free atmosphere. We also consider a model variant that assigns independent elemental abundances to the dayside and nightside while retaining local chemical equilibrium within each hemisphere. This prescription provides a phenomenological representation of hemispheric chemical differences, motivated by the enhanced nightside CH$_4$ abundance reported by \mbox{\citet{Evans-Soma+etal+2025}}, without explicitly modelling chemical kinetics and atmospheric transport. Cloud effects are assessed independently using the simplified grey-cloud prescription described in Section~\ref{RETR}. Although condensates are unlikely to remain stable over the hottest dayside regions, they may form on the cooler nightside and limbs \citep[e.g.,][]{Parmentier+etal+2016,Parmentier+etal+2018b,Parmentier+etal+2021,Komacek+etal+2022,Helling+etal+2023}. Their contribution to the thermal emission spectrum is nevertheless expected to be limited across the NIRSpec/G395H bandpass \mbox{\citep{Evans-Soma+etal+2025}}.

Future iterations of this framework could incorporate more complex atmospheric processes. While fully self-consistent chemical kinetics (including disequilibrium chemistry) and dynamical cloud models are computationally prohibitive for retrieval, we propose extending the physically motivated parameterisation approach used here for temperature to these other domains. This contrasts with ``free'' parameterisations. For instance, \citet{Irwin+etal+2020} employs prescribed meridional functions together with interpolation at selected points on the grid. This data-driven formulation relaxes physical constraints but introduces a large number of free parameters and substantial, often difficult-to-quantify, degeneracies. Instead, we advocate for deriving analytically tractable expressions from simplified governing equations (e.g., chemical kinetics or cloud microphysics) via nondimensionalisation. Although developing such simplified yet physically robust models for chemistry and clouds remains challenging, it offers a promising pathway to interpret high-quality data from \textit{JWST} and \textit{Ariel} with minimal arbitrary assumptions. The development of such unified physically driven frameworks is underway.

\subsection{Radiative Transfer in Multidimensional Atmospheres}
\label{sec:rt_multid_phase}

With the 3D temperature field $T(\theta,\phi,p)$ specified (see Section~\ref{sec:para_tp}), our forward model proceeds in three stages: (i) computing the local emergent intensity for each atmospheric column using a 1D radiative transfer code; (ii) integrating these intensities over the visible planetary disk at each orbital phase to generate a phase-wavelength flux cube; and (iii) projecting this cube into the observational frame for direct comparison with data. Throughout this work, we assume the planet is tidally locked (i.e., synchronous rotation), with the substellar point fixed at longitude $\theta=0$ in the corotating frame.

We discretise the planetary atmosphere into a uniform grid of $N_{\theta}\times N_{\phi}\times N_{p}$ points. For each atmospheric vertical column $(\theta_i,\phi_j)$, we compute the emergent thermal emission spectrum under the assumption of local thermodynamic equilibrium (LTE). We utilise \textit{petitRADTRANS} in emission geometry, neglecting scattering for computational efficiency. Line opacities are computed in the line-by-line mode using high-resolution opacity tables generated with \texttt{pyROX} \citep{deRegt+etal+2025} and formatted for \texttt{petitRADTRANS 3}, which include pressure broadening, while continuum sources (i.e., H$_2$-H$_2$ and H$_2$-He CIA, Rayleigh scattering by H$_2$ and He, and H$^-$ bound-free and free-free absorption) are included consistently. The emergent intensity is evaluated on a common wavelength grid, with spectra computed using the line-by-line method and resampled to the requested resolving power. The resulting per-column spectra, $I_{i,j,\ell}$, with isotropic assumption, where $\ell$ indexes the wavelength grid, provide the input for disk integration.

To obtain the disk-integrated flux at an orbital phase angle $\varphi$ (in the tidally locked frame), we integrate the emergent specific intensities from all atmospheric columns over the visible hemisphere weighted by their projected area. A surface element at $(\theta,\phi)$ is visible to the observer if $\cos\vartheta = \cos\phi\,\cos(\theta+\varphi) > 0$, where $\vartheta$ is the emission angle between the local surface normal and the observer's line of sight. The disk-integrated flux is: 
\begin{equation}
F_{\rm p}(\varphi,\lambda)
=
\iint_{\cos\vartheta>0}
I(\theta,\phi,\lambda)\,
\cos\vartheta\,\cos\phi\,
R_{\rm p}^2\,{\rm d}\phi\,{\rm d}\theta.
\end{equation}
With $\cos\vartheta = \cos\phi\cos(\theta+\varphi)$, this reduces to
\begin{equation}
F_{\rm p}(\varphi,\lambda)
=
\iint_{\cos\vartheta>0}
I(\theta,\phi,\lambda)\,
\cos^2\phi\,\cos(\theta+\varphi)\,
R_{\rm p}^2\,{\rm d}\phi\,{\rm d}\theta,
\end{equation}
where $I(\theta,\phi, \lambda)$ is the top-of-atmosphere specific intensity (wavelength space), $\lambda$ is the wavelength, and $R_{\rm p}$ is the planetary radius \citep[equivalent to the formulation in ][]{Morris+etal+2022,Cowan+etal+2011}. In practice, the integral is evaluated on a discrete latitude–longitude grid, with north–south symmetry applied for computational efficiency, as the adopted $\cos\phi$ latitudinal dependence is symmetric about the equator. This formulation naturally accounts for the changing visibility of longitudes as the planet orbits. Near the terminator, partially visible longitude cells are treated by analytically integrating the overlap with the visible region, which avoids aliasing and guarantees flux conservation. Repeating this process for all orbital phases $\varphi_m$ (where $m$ indexes the phase angle grid) yields the phase-wavelength cube, $\{F(\varphi_m, \lambda_\ell)\}$, to use in the retrieval.

To map the theoretical model to the observed data space, we apply instrumental broadening, normalisation by the stellar flux, and Doppler shifts induced by the planet's orbital motion. For the stellar reference, we adopt synthetic stellar atmosphere spectra from the PHOENIX library, accessed via the \texttt{pysynphot} package \citep{STScIDevelopmentTeam+etal+2013}, and interpolated to the stellar parameters. The final model output is the phase-resolved flux ratio, $1+F_{\rm p}/F_{\rm s}$, resampled onto the detector's wavelength grid, to compare with the observed residual spectral matrix for likelihood evaluation.

\begin{figure}
\centering
\includegraphics[width=0.45\textwidth, trim=10 10 10 0]{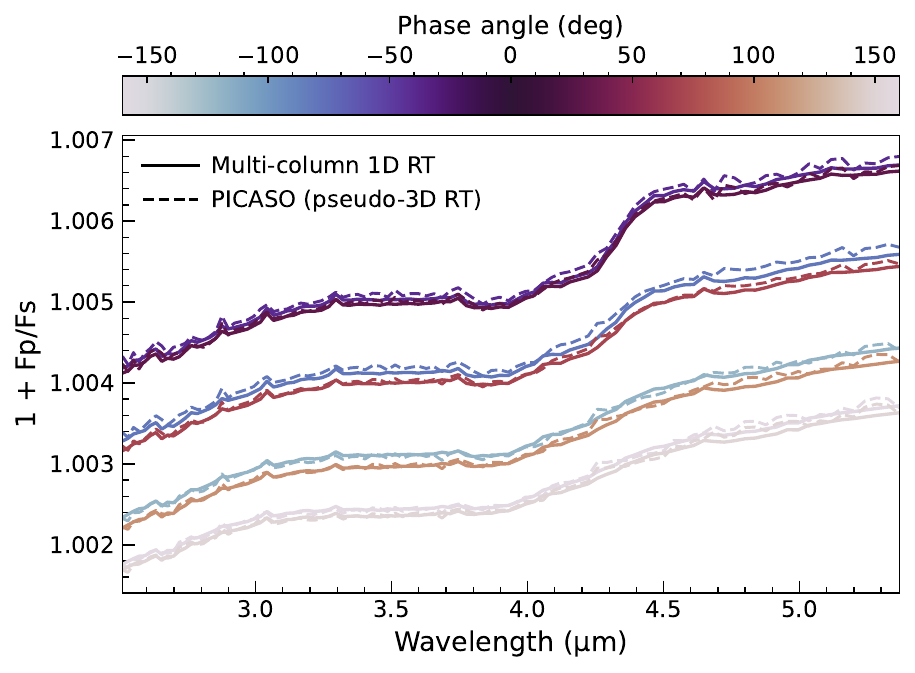}
\caption{Comparison between phase-resolved emission spectra computed using the pseudo-3D code PICASO (dashed lines) and those obtained from our disk-integrating multiple 1D radiative-transfer framework (solid lines). Different colours correspond to spectra at various orbital phases.}
\label{Fig7}
\end{figure}

Compared with phase-curve spectroscopy methods that rely on full 3D calculations, the rapid approach presented here can be efficiently incorporated into atmospheric retrieval frameworks, thereby enabling multidimensional structural retrievals. Figure~\ref{Fig7} compares a spectroscopic phase curve computed using our rapid approach, which employs 1D plane-parallel radiative transfer, against one generated using \textit{PICASO} \citep[v3.1.2,][]{Batalha+etal+2019, Mukherjee+etal+2023}, which adopts a plane-parallel radiative transfer framework but treats the angle-dependent emergent intensity in a more realistic manner during disk integration. We note that fully self-consistent 3D radiative transfer calculations require dedicated solvers, such as \textit{gCMCRT} \citep{Lee+etal+2022}, and are computationally prohibitive for iterative retrieval applications. The differences between the two approaches are limited to the tens-to-hundred ppm level across all phases. Given that the measurement uncertainties of individual JWST spectra are typically on the same order, our rapid approach provides sufficient accuracy for retrieving large-scale atmospheric structures while enabling the computational speed required for iterative retrieval frameworks.

\section{Retrieval Analyses}\label{RETR}

\subsection{Bayesian Inference Framework}

To infer the atmospheric properties of WASP-121b, we perform retrievals on the phase-resolved spectra binned over 10 detector pixels per wavelength channel. The core of the analysis involves comparing the observed residual data matrix, $R(\varphi, \lambda)$ (derived in Section~\ref{sec:data_ana}), against a forward model, $M(\varphi,\lambda)$, which incorporates both the astrophysical signal and instrumental systematics. The general form of the forward model is: 
\begin{equation}
M(\varphi,\lambda) = \mathcal{F}(\varphi,\lambda) \times \gamma(\varphi,\lambda),
\end{equation}
where $\mathcal{F}$ represents the astrophysical component and $\gamma$ is the instrumental systematics term. If the instrumental systematics have already been corrected during the eclipse normalisation process (as described in Section \ref{sec:data_ana}), then $\gamma=1$, and the forward model simplifies to the astrophysical term alone. In this work, rather than parameterising instrumental systematics with complex functions during the retrieval, we leverage the two secondary eclipses to linearly detrend the data beforehand. As a result, the retrieval focuses solely on modelling the astrophysical signal. The astrophysical component is defined as: 
\begin{equation}
\mathcal{F}(\varphi,\lambda) \;=\; 1+\frac{F_{\rm p}(\varphi,\lambda)}{F_{\rm s}(\lambda)},
\end{equation}
where the planetary flux $F_{\rm p}$ is computed via the method detailed in Section~\ref{sec:rt_multid_phase}. The stellar reference spectrum, $F_{\rm s}(\lambda)$, is generated from the PHOENIX library using \texttt{pysynphot}, interpolated to the specific stellar parameters of WASP-121. 

Assuming a standard Gaussian log-likelihood: 
\begin{equation}\label{Likefun}
\ln L=-\frac{1}{2} \sum_{i, j}\left[\frac{\left(R_{i j}-M_{i j}\right)^{2}}{\left(\beta \sigma_{i j}\right)^{2}}+\ln 2 \pi\left(\beta \sigma_{i j}\right)^{2}\right]
\end{equation}
where \(\sigma_{ij}\) are the propagated uncertainties, and \(\beta\) is a global scale factor of the uncertainties. This likelihood function is commonly adopted in high-resolution spectroscopic retrieval analyses \citep[e.g.,][]{Brogi+etal+2019,Gibson+etal+2020,Yan+etal+2020,Yang+etal+2025}. 
 We explore the parameter space and compute the posterior distributions using the deep-learning-assisted nested sampling algorithm \texttt{NAUTILUS} \citep{Lange+2023}, which provides substantially improved sampling efficiency for high-dimensional retrieval problems. A comparison between \texttt{NAUTILUS} and the traditional \texttt{MultiNest}\citep{Skilling+etal+2004,Feroz+etal+2009} sampler for the present retrieval framework is provided in Appendix~\ref{app_A}. The retrievals were performed with $N_{\rm live}=1000$, a target effective sample size of $N_{\rm eff}=10000$, and a termination threshold of $f_{\rm live}=0.1$.

\subsection{Model Configuration and Priors}

We carried out a suite of multidimensional retrieval analyses for WASP-121b using the JWST/NIRSpec G395H phase-curve observations. 
All retrieval models share the same temperature parameterisation, radiative-transfer framework, and Bayesian inference setup. The model variants differ in their chemical treatment, cloud prescription, treatment of nightside thermal inversions, and allowance for an additional offset between the NRS1 and NRS2 detectors. The chemical variants assume local equilibrium, with elemental abundances either shared globally or retrieved independently for the dayside and nightside.
The retrieval framework is summarised below, with all free parameters and their priors listed in Table~\mbox{\ref{Tab1}}. The upper bound of the temperature range is set by the validity range of the opacity tables adopted in the radiative transfer calculations (i.e., 4500 K).

For the temperature field, the thermal structure is parameterised using the physically motivated framework described in Section~\ref{sec:para_tp}. We adopt five pressure-dependent control functions, \(T_n\), \(T_d\), \(\varepsilon\), \(\psi\), and \(\alpha\), together with the pressure-independent exponent \(\gamma_{\rm lat}\). For the retrieval, we replace the prescribed \(\cos\phi\) latitudinal factor in Equation~\ref{ther_eq4} with \(\cos^{\gamma_{\rm lat}}\!\phi\), allowing the phase-curve data to constrain the meridional extent of the irradiation-driven temperature contrast rather than imposing a fixed cosine profile. The original form is recovered for \(\gamma_{\rm lat}=1\), whereas values below or above unity produce a broader or more equatorially concentrated temperature distribution, respectively. All latitudes share the same pressure-dependent \(T_n\), \(T_d\), \(\alpha\), \(\varepsilon\), and \(\psi\).
In the \textsc{Base} retrieval, the reference nightside $T_{n}(p)$ and dayside vertical temperature profiles $T_{d}(p)$ are described by the analytic irradiated-atmosphere profile of \citet{Line+etal+2013}. Separate parameter sets are used for the $T_{n}(p)$ and $T_{d}(p)$, each containing the infrared opacity, two visible-to-infrared opacity ratios, and the irradiation partitioning parameters, namely $\log\kappa_{\rm IR}, \log\gamma, \log\gamma_{2}, T_\alpha, T_\beta$. The internal temperature is fixed to $T_{\rm int}=500$~K, while the equilibrium temperature is computed from the retrieved stellar effective temperature, with the stellar radius and orbital separation fixed to the system values \mbox{\citep{Evans-Soma+etal+2025}}. 
The longitudinal redistribution parameters $\varepsilon$ and $\psi$ are parameterised as log-linear functions of pressure,
\begin{equation}
\log_{10}\varepsilon(p)=\varepsilon_{0}+\varepsilon_{1}\zeta(p), \qquad
\log_{10}\psi(p)=\psi_{0}+\psi_{1}\zeta(p),
\end{equation}
where $\zeta(p)$ denotes a dimensionless logarithmic pressure coordinate,
\begin{equation}
\zeta(p)=2\,\frac{\log_{10}(p/p_{\rm min})}
{\log_{10}(p_{\rm max}/p_{\rm min})}-1 .
\end{equation}
This mapping transforms the pressure interval $p_{\rm min}\leq p\leq p_{\rm max}$ onto $-1\leq\zeta\leq1$. In the retrieval, we adopt $p_{\rm min}=10^{-5}$~bar and $p_{\rm max}=10^{1}$~bar. The $\alpha$ is likewise allowed to vary with pressure and is represented by a logistic function,
\begin{equation}
\alpha(p)=\left[1+\exp\left(-\frac{\zeta(p)-\zeta_{\alpha}}{\Delta\zeta_{\alpha}}\right)\right]^{-1},
\end{equation}
which enforces $0\leq\alpha(p)\leq1$.

For chemistry and opacities, we assume chemical equilibrium and a cloud-free atmosphere for the \textsc{Base} retrieval, following \mbox{\citet{Evans-Soma+etal+2025}} for WASP-121b. The chemical composition is governed by four free parameters: the specific elemental abundances [O/H], [C/H], and [Si/H], and a bulk metallicity [M/H] scaling all other elements.  We include line opacities for key species in hot Jupiters that are expected to be abundant under chemical equilibrium and to contribute significantly to the opacity budget within the NIRSpec/G395H wavelength range, including CO \citep{Li+etal+2015}, H$_2$O \citep{Rothman+etal+2010}, CO$_2$ \citep{Yurchenko+etal+2020}, NH$_3$ \citep{Coles+etal+2019}, CH$_4$ \citep{Hargreaves+etal+2020}, C$_2$H$_2$ \citep{Chubb+etal+2020}, SiO \citep{Yurchenko+etal+2022}, HCN \citep{Barber+etal+2014,Harris+etal+2006}, FeH \citep{Bernath+etal+2020}, TiO \citep{McKemmish+etal+2019}, and VO \citep{McKemmish+etal+2016}. The line opacities are generated with \texttt{pyROX}, rather than taken directly from the default precomputed \textit{petitRADTRANS} opacity database. While recent studies suggest nightside disequilibrium chemistry may be relevant \mbox{\citep{Evans-Soma+etal+2025}}, incorporating chemical kinetic transport or multidimensional chemical/cloud parameterisations is beyond the scope of this initial study and is reserved for future work. 

For the phase-resolved thermal emission radiative transfer calculations, we discretise the planetary atmosphere into a grid of $N_{\phi}=5$ latitudes, $N_{\theta}=12$ longitudes, and $N_{\rm p}=31$ pressure layers. This resolution provides an appropriate balance between resolving the large-scale phase variation and maintaining computational feasibility for Bayesian retrieval.

In addition to the \textsc{Base} retrieval, we explored a small set of controlled model variants to assess the robustness of the inferred atmospheric structure. The cloud-including variant (\textsc{Cloud}) adopts a minimal three-parameter pseudo-condensation prescription. Rather than modelling detailed cloud microphysics, particle-size distributions, or scattering properties, this prescription assigns each atmospheric column an effective grey opaque cloud-top pressure \(P_{\rm cloud}\), which is passed directly to \texttt{petitRADTRANS}. The pseudo-condensation curve is
\begin{equation}
T_{\rm cond}(p)=T_{\rm cond,ref}
+s_{\rm cond}\log_{10}\left(\frac{p}{p_{\rm ref}}\right),
\end{equation}
with \(p_{\rm ref}=1~{\rm bar}\). Layers colder than \(T_{\rm cond}(p)\) are classified as cloudy. The cloud base \(P_{\rm base}\), is defined as the deepest cloudy layer, and the cloud top is placed at
\begin{equation}
P_{\rm cloud}=P_{\rm base}\,10^{-\Delta\log P_{\rm cloud}} .
\end{equation}
The \(T_{\rm cond,ref}\), \(s_{\rm cond}\), and \(\Delta\log P_{\rm cloud}\) are treated as free parameters. This grey-cloud model therefore provides a conservative test of whether the observations require optically thick clouds over part of the planet.

As a second variant (\textsc{NoInv}), we imposed a monotonic nightside temperature structure to test the role of nightside thermal inversions. For all atmospheric columns on the nightside, defined by longitude \(|\theta|\geq\pi/2\), the temperature is required to decrease monotonically with decreasing pressure over \(10^{-5}\)--\(10^{1}\) bar. Any proposed sample that violates this condition is rejected during this retrieval. This variant tests whether the observations favour a nightside thermal inversion.

As a third variant (\textsc{Offset}), we allowed for a relative flux-calibration offset between NRS1 and NRS2. NRS1 is used as the reference channel, and a single additive flux offset is applied uniformly to all NRS2 bins at all orbital phases. The offset is wavelength independent and affects only the NRS2 flux zero point.

As a fourth variant (\textsc{DNChem}), we allowed the dayside and nightside elemental abundances to vary independently, while retaining local chemical equilibrium within each hemisphere. Specifically, separate values of [O/H], [C/H], [Si/H], and [M/H] were assigned to the dayside and nightside atmospheric columns.

\begin{table*}[htb!]
\caption{
Free parameters and priors used in the retrievals. Model labels denote the baseline cloud-free model with globally shared elemental abundances (\textsc{Base}), the model with independent dayside and nightside elemental abundances (\textsc{DNChem}), the grey-cloud model (\textsc{Cloud}), the nightside no-inversion model (\textsc{NoInv}), and the NRS2-offset model (\textsc{Offset}).
The elemental-abundance parameters are expressed in dex relative to the solar abundance scale adopted by \textit{FastChem}.
}
\label{Tab1}
\centering
\begin{tabular}{l l c c}
\hline\hline
Parameter & Description [Unit] & Prior & Models \\
\hline
\multicolumn{4}{l}{\textbf{Reference temperature profiles}} \\
\hline
$\log\kappa_{{\rm IR},n,d}$ & IR opacity & $\mathcal{U}(-5.0,-0.5)$ & All \\
$\log\gamma_{n,d}$ & Visible/IR ratio & $\mathcal{U}(-4.0,1.5)$ & All \\
$\log\gamma_{2,n,d}$ & Second visible/IR ratio & $\mathcal{U}(-4.0,1.5)$ & All \\
$T_{\alpha,n,d}$ & Visible channels partition & $\mathcal{U}(0,1)$ & All \\
$T_{\beta,n,d}$ & Redistribution factor & $\mathcal{U}(0,2)$ & All \\
$\gamma_{\rm lat}$ & Latitudinal exponent & $\mathcal{U}(0,2)$ & All \\
\hline
\multicolumn{4}{l}{\textbf{Longitudinal redistribution}} \\
\hline
$\varepsilon_{0}$ & $\log_{10}\varepsilon$ intercept & $\mathcal{U}(-2.2,0.8)$ & All \\
$\varepsilon_{1}$ & $\log_{10}\varepsilon$ slope & $\mathcal{U}(-1.5,1.5)$ & All \\
$\psi_{0}$ & $\log_{10}\psi$ intercept & $\mathcal{U}(-2.2,0.8)$ & All \\
$\psi_{1}$ & $\log_{10}\psi$ slope & $\mathcal{U}(-1.5,1.5)$ & All \\
$\zeta_{\alpha}$ & $\alpha$ transition location & $\mathcal{U}(-2.0,0.5)$ & All \\
$\Delta\zeta_{\alpha}$ & $\alpha$ transition width & $\mathcal{U}(0.2,2.5)$ & All \\
\hline
\multicolumn{4}{l}{\textbf{Elemental abundances}} \\
\hline
$\mathrm{[O/H]}$ & O abundance [dex] & $\mathcal{U}(0.5,1.7)$ & All except \textsc{DNChem} \\
$\mathrm{[C/H]}$ & C abundance [dex] & $\mathcal{U}(0.5,1.7)$ & All except \textsc{DNChem} \\
$\mathrm{[Si/H]}$ & Si abundance [dex] & $\mathcal{U}(0.5,1.7)$ & All except \textsc{DNChem} \\
$\mathrm{[M/H]}$ & Bulk metallicity [dex] & $\mathcal{U}(-1,1)$ & All except \textsc{DNChem} \\
$\mathrm{[O/H]}_{d,n}$ & Dayside/nightside O abundance [dex] & $\mathcal{U}(0.5,1.7)$ & \textsc{DNChem} \\
$\mathrm{[C/H]}_{d,n}$ & Dayside/nightside C abundance [dex] & $\mathcal{U}(0.5,1.7)$ & \textsc{DNChem} \\
$\mathrm{[Si/H]}_{d,n}$ & Dayside/nightside Si abundance [dex] & $\mathcal{U}(0.5,1.7)$ & \textsc{DNChem} \\
$\mathrm{[M/H]}_{d,n}$ & Dayside/nightside bulk metallicity [dex] & $\mathcal{U}(-1,1)$ & \textsc{DNChem} \\
\hline
\multicolumn{4}{l}{\textbf{Orbital, stellar, and noise parameters}} \\
\hline
$\Delta v$ & Velocity shift [km\,s$^{-1}$] & $\mathcal{U}(-150,150)$ & All \\
$K_{\rm p}$ & Orbital velocity [km\,s$^{-1}$] & $\mathcal{U}(210,225)$ & All \\
$\beta_{\rm scale}$ & Error scaling & $\mathcal{U}(0.1,5)$ & All \\
$T_{\rm eff}$ & Stellar effective temperature [K] & $\mathcal{U}(6400,6800)$ & All \\
\hline
\multicolumn{4}{l}{\textbf{Model-specific parameters and constraints}} \\
\hline
$T_{\rm cond,ref}$ & Condensation reference temperature $T$ [K] & $\mathcal{U}(1000,3000)$ & \textsc{Cloud} \\
$s_{\rm cond}$ & Condensation temperature gradient [K\,dex$^{-1}$] & $\mathcal{U}(0,400)$ & \textsc{Cloud} \\
$\Delta\log P_{\rm cloud}$ & Cloud extent [dex] & $\mathcal{U}(0,4)$ & \textsc{Cloud} \\
Nightside monotonicity & No inversion & Rejection constraint & \textsc{NoInv} \\
$\Delta_{\rm NRS2}$ & NRS2 flux offset [ppm] & $\mathcal{U}(-3000,3000)$ & \textsc{Offset} \\
\hline
\end{tabular}
\end{table*}

\subsection{Results: Retrieved Atmospheric Structures}

The Bayesian evidences of the retrieval models are summarised in Table~\ref{Tab2}. The \textsc{DNChem} model yields the largest evidence, with \(\ln\mathcal{Z}=89711.96\), exceeding the evidence of \textsc{Base} by \(\Delta\ln\mathcal{Z}=65.31\). The corresponding corner plots are shown in Figures~\ref{FigA5} and Figures~\ref{FigA6}, respectively. The data therefore strongly favour the model that allows independent dayside and nightside elemental abundances within the local chemical equilibrium framework. This preference is also compatible with the enhanced nightside CH$_4$ abundance reported by \mbox{\citet{Evans-Soma+etal+2025}}, which they interpreted as evidence for disequilibrium chemistry.
The specific grey-cloud prescription remains disfavoured, although this does not exclude more localised or microphysically complex clouds.
In contrast, the substantially lower evidence of the \textsc{NoInv} model indicates that a strictly monotonic nightside temperature profile is disfavoured, implying the presence of a thermal inversion in at least part of the nightside atmosphere. The \textsc{Offset} model is also strongly disfavoured, indicating that the retrieved atmospheric signal is unlikely to arise from an unaccounted flux offset between the NRS1 and NRS2 detectors.

\begin{table}[htb!]
\caption{
Bayesian evidence comparison for the retrieval models. The evidence difference is computed relative to the baseline cloud-free model,
\(\Delta\ln\mathcal{Z}=\ln\mathcal{Z}-\ln\mathcal{Z}_{\rm Base}\).
}
\label{Tab2}
\centering
\begin{tabular}{l l c c}
\hline\hline
Model & Description & $\ln\mathcal{Z}$ & $\Delta\ln\mathcal{Z}$ \\
\hline
\textsc{Base} 
& Clear, global abundances 
& 89646.65 
& -- \\

\textsc{DNChem}
& Day/night abundances
& 89711.96
& $+65.31$ \\

\textsc{Cloud} 
& Grey-cloud 
& 89629.60 
& $-17.05$ \\

\textsc{NoInv} 
& No nightside inversion
& 88925.37 
& $-721.28$ \\

\textsc{Offset} 
& NRS1-NRS2 flux-offset 
& 88476.48 
& $-1170.17$ \\
\hline
\end{tabular}
\end{table}

\begin{figure*}[htb!]
\centering
\includegraphics[width=\textwidth]{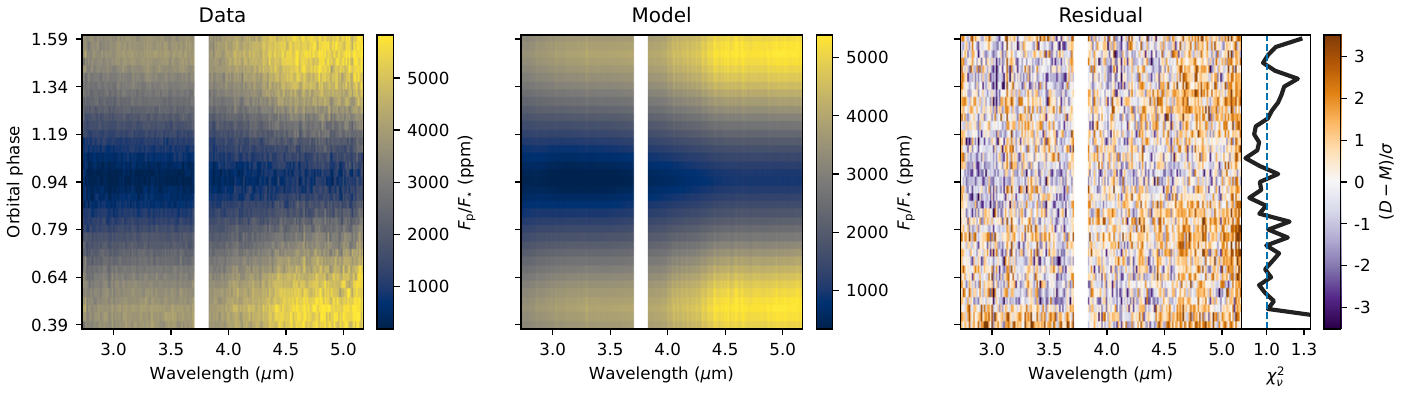}
\caption{
Phase-resolved fit of the \textsc{DNChem} retrieval to the JWST/NIRSpec G395H emission spectra of WASP-121b. 
The left and middle panels show the observed planet-to-star flux ratio (excluding the transit and secondary eclipse) and the best-fitting model, respectively, as functions of wavelength and orbital phase. 
The right panel shows the residuals normalised by the observational uncertainties \((D-M)/\sigma\), together with the phase-dependent \(\chi^2_\nu\). 
}
\label{Fig8}
\end{figure*}

Therefore, we focus on \textsc{DNChem}, the highest-evidence model, as the reference retrieval for the following analysis.
Figure~\ref{Fig8} compares the eclipse-normalised spectroscopic phase-curve data with the best-fitting model and corresponding residuals. The model successfully reproduces the observed two-dimensional phase--wavelength structure across the NIRSpec/G395H bandpass, with residuals that are largely consistent with the noise level and phase-dependent $\chi^2_\nu$ values close to unity. An alternative visualisation of the model fit is provided by the one-dimensional spectra at representative orbital phases shown in Figure~\ref{FigA2}. Nevertheless, localised residuals remain on the nightside, particularly at orbital phases $\phi \lesssim 0.25$ and $\phi \gtrsim 0.75$, and at wavelengths longer than $\sim4.5~\mu{\rm m}$. These discrepancies may indicate additional atmospheric complexity not captured by the model, including complex cloud structures or chemistry, although residual instrumental systematics and temporal atmospheric variability cannot be excluded. Such features motivate future retrievals with more flexible atmospheric parameterisations.

\begin{figure*}[htbp]
\centering
\includegraphics[width=\textwidth]{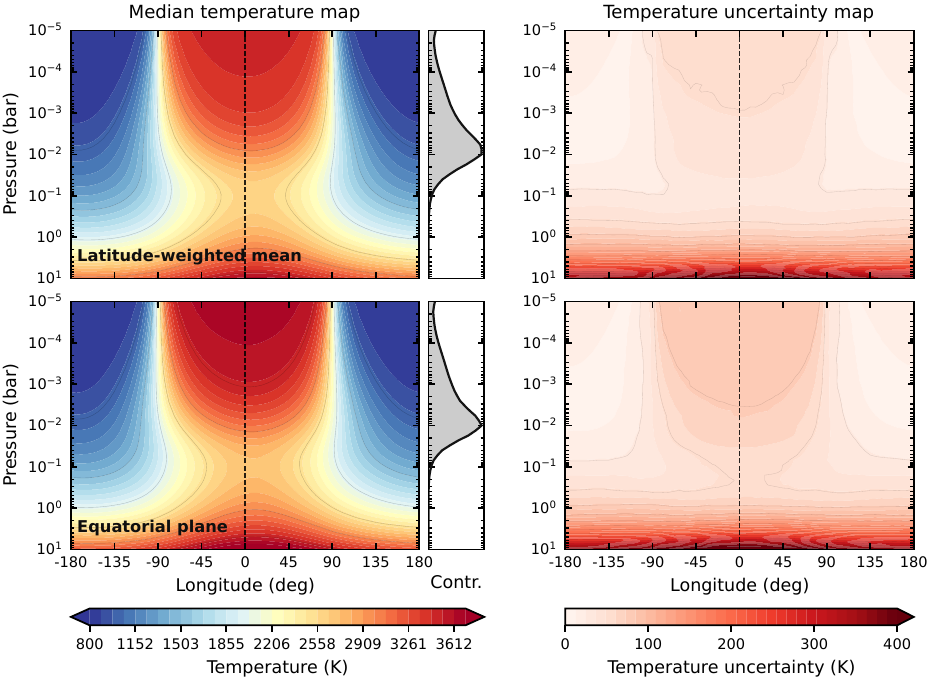}
\caption{
Longitude-pressure temperature structure retrieved from the \textsc{DNChem} model.
The left column shows the posterior median temperature field and the right column shows the corresponding \(1\sigma\) posterior uncertainty.
The upper row gives the latitude-weighted mean temperature field, while the lower row shows the equatorial plane.
The grey profiles beside the temperature maps show the normalised G395H contribution functions.
The dashed vertical line marks the substellar longitude.
The retrieved atmosphere shows a strong day--night temperature contrast and a dayside thermal inversion at low pressures.
}
\label{Fig9}
\end{figure*}

\begin{figure*}[htbp]
\centering
\includegraphics[width=\textwidth]{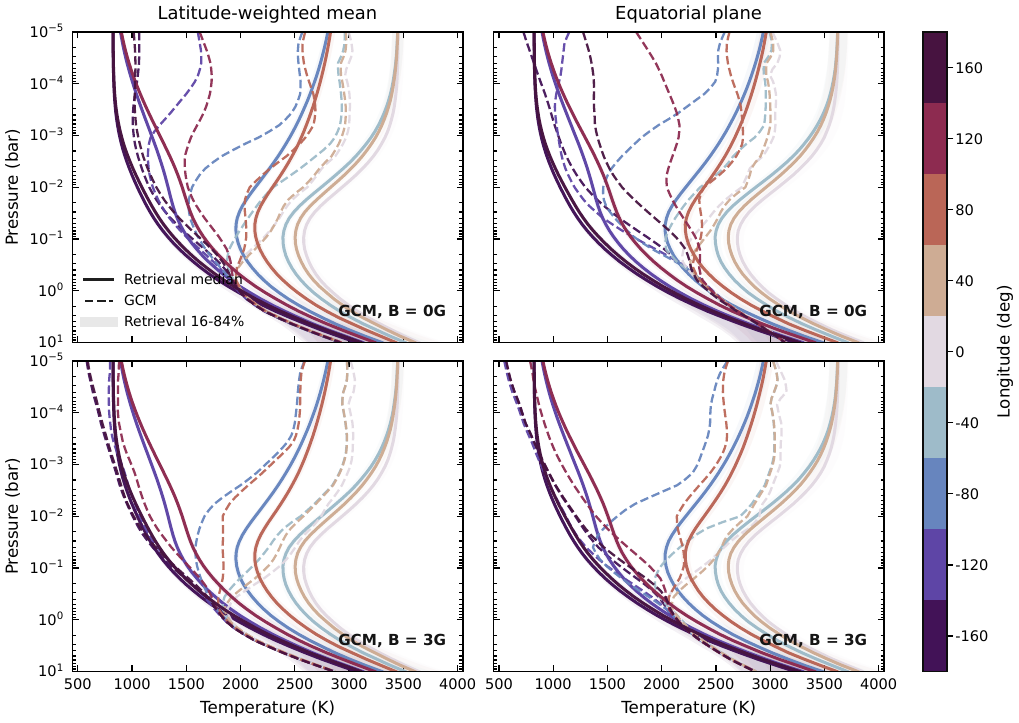}
\caption{
Longitude-resolved temperature profiles from the \textsc{DNChem} retrieval compared with representative GCM temperature profiles.
The upper and lower rows compare the retrieval with GCM simulations assuming planetary magnetic-field strengths of $0$ and $3$~G, respectively.
In each row, the left panel shows latitude-weighted mean profiles and the right panel shows equatorial-plane profiles.
Solid curves denote the retrieved posterior median profiles, shaded bands show the $1\sigma$ uncertainties, and dashed curves show the corresponding GCM profiles.
Colours indicate longitude.
}
\label{Fig10}
\end{figure*}

The retrieved longitude--pressure temperature distributions from the \textsc{DNChem} model are shown in Figure~\ref{Fig9}. The corresponding longitude-resolved profiles are compared in Figure~\ref{Fig10} with the WASP-121b SPARC/MITgcm GCM simulations of \citet{Fecanin+etal+2026}, which incorporate spatially varying magnetic drag within the ADAM framework. The corresponding pressure-wavelength and pressure-longitude contribution functions are presented in Appendix Figure~\ref{FigA3}. The retrieved atmosphere exhibits a strong longitudinal dichotomy. The dayside remains hot over a broad range of pressures, with temperatures exceeding $\sim3000~{\rm K}$ in the upper atmosphere, whereas the nightside is substantially cooler throughout the pressure range. The dayside hot region extends from the substellar point to the terminators, while longitudinal temperature variations within each hemisphere remain modest relative to the vertical structure. At greater pressures, the day--night temperature contrast diminishes as the profiles converge. 

Figure~\ref{Fig10} compares the retrieved longitude-resolved temperature profiles with a non-magnetic GCM ($0$~G) and a GCM incorporating magnetic drag associated with a $3$~G planetary magnetic field that has no dipole tilt. The retrieved profiles exhibit a pronounced dayside thermal inversion that weakens towards the antistellar hemisphere, where the profile becomes non-inverted. Thermal inversions persist near both limbs, with the eastern and western limb temperatures differing by several hundred kelvin over the pressures probed by G395H. Relative to the $0$~G GCM, the retrieved temperature field shows a more longitudinally confined dayside hot region and a larger day--night temperature contrast. Its pressure–temperature structure therefore more closely resembles the magnetic \(3\)~G GCM case, although similar behaviour may also arise from other damping mechanisms, such as Rayleigh drag. Further constraints are needed to distinguish between these possibilities. 

\begin{figure}[htbp]
\centering
\includegraphics[width=0.48\textwidth]{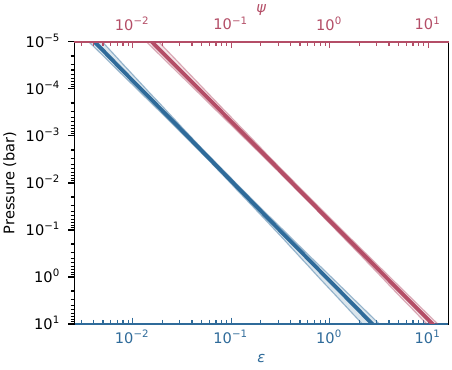}
\caption{
Retrieved vertical profiles of the dimensionless transport parameters $\varepsilon(p)$ and $\psi(p)$ from the \textsc{DNChem} model.
The bottom blue axis shows $\varepsilon$, and the top red axis shows $\psi$.
Solid curves denote posterior medians, while shaded regions and thin curves show the $1\sigma$ uncertainties.
Both parameters increase monotonically with pressure, indicating that dynamical heat redistribution becomes progressively more efficient at depth.
}
\label{Fig11}
\end{figure}

The retrieved vertical profiles of the dimensionless transport parameters $\varepsilon(p)$ and $\psi(p)$ are shown in Figure~\ref{Fig11}. Both parameters increase monotonically with pressure. Across $10^{-5}$--10~bar, the posterior medians of $\varepsilon$ and $\psi$ increase from $0.004$ to $2.710$ and from $0.016$ to $11.110$, respectively. This behaviour indicates that radiative relaxation is faster than both advective and diffusive transport at low pressures, whereas values above unity at depth imply that dynamical redistribution becomes increasingly effective relative to radiation. This pressure dependence is consistent with the reduced day--night temperature contrast at depth in Figure~\ref{Fig9}.

\begin{figure}[htbp]
\centering
\includegraphics[width=0.48\textwidth]{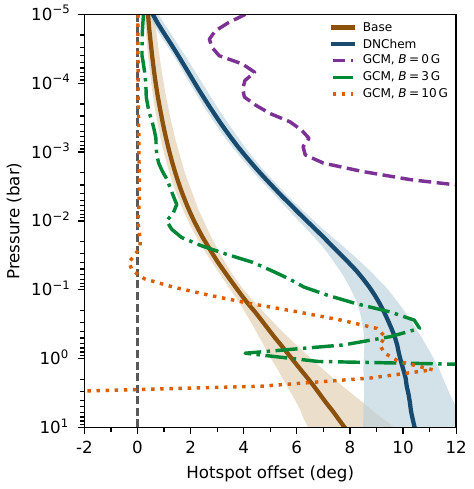}
\caption{
Pressure-dependent hotspot offsets inferred from the \textsc{Base} and \textsc{DNChem} retrievals.
The hotspot longitude is defined as the longitude of the maximum latitude-weighted temperature at each pressure level.
Solid curves show the posterior medians, and shaded bands denote the corresponding $1\sigma$ uncertainties.
Coloured curves show the corresponding hotspot offsets from representative GCM cases with different values of the dipole magnetic field strength \(B\).
}
\label{Fig12}
\end{figure}

The pressure-dependent hotspot offset provides an independent view of how the retrieved temperature field redistributes stellar energy (see Figure~\ref{Fig12}). We define the hotspot longitude at each pressure as the longitude of the maximum latitude-weighted temperature. The \textsc{DNChem} and \textsc{Base} retrievals place the hotspot close to the substellar point in the upper atmosphere, although \textsc{DNChem} yields systematically larger offsets than \textsc{Base}. Across the pressures probed primarily by G395H, the offset of the hotspot increases from $4.03^{\circ}$ near $10^{-3}$~bar to $8.73^{\circ}$ near $0.1$~bar for \textsc{DNChem}, compared to an increase from $1.23^{\circ}$ to $3.69^{\circ}$ for \textsc{Base}. Thus, both retrievals favour a coherent pressure-dependent displacement of the thermal maximum rather than a large photospheric hotspot shift, with \textsc{DNChem} supporting a displacement stronger than \textsc{Base}. This behaviour is consistent with the temperature maps in Figure~\ref{Fig9}: rapid radiative relaxation and/or strong drag can maintain the upper-atmosphere hotspot near the substellar point, whereas increasing radiative-to-dynamical timescale ratios at depth permit a larger longitudinal displacement \citep{Zhang+etal+2017}. Across the pressures most relevant to G395H, both retrievals yield smaller offsets than the non-magnetic GCM. Within the two magnetic-field cases considered here, the \textsc{Base} offsets lie closer to the $3$~G prediction, whereas the \textsc{DNChem} offsets generally lie between the $0$ and $3$~G predictions. The underlying drag mechanism cannot be uniquely determined because alternatives such as Rayleigh drag cannot be excluded.

\begin{figure}[htbp]
\centering
\includegraphics[width=\columnwidth]{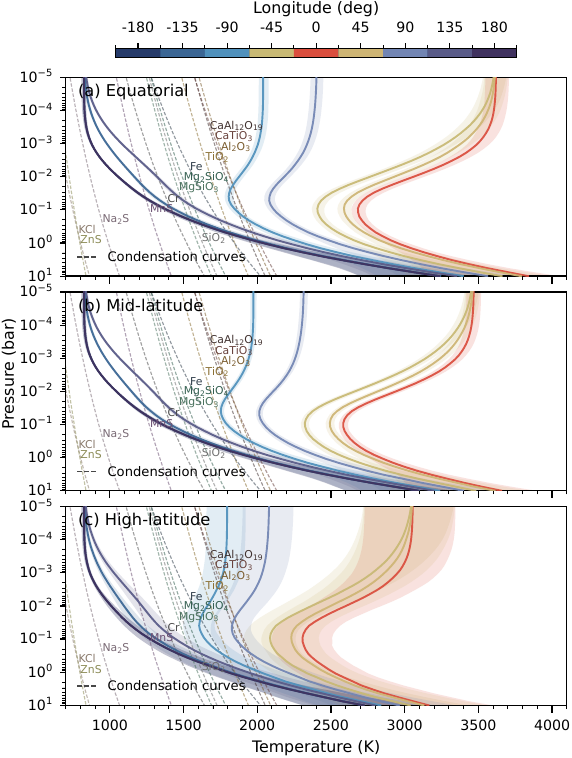}
\caption{
Retrieved temperature profiles compared with equilibrium condensation curves.
Panels (a)--(c) show the equatorial, mid-latitude, and high-latitude profiles, respectively. Within each panel, solid coloured curves show the posterior medians from the \textsc{DNChem} retrieval at different longitudes.
The colour bar indicates longitude.
Shaded bands show the $1\sigma$ uncertainties of the retrieved profiles.
Labelled dashed curves show representative condensation curves for candidate condensate species calculated assuming solar metallicity.
The hottest dayside profiles remain above most condensation curves, whereas the nightside, limb, and high-latitude profiles intersect several refractory condensation curves, indicating that condensates could be thermally stable only in restricted regions of the three-dimensional atmosphere.
}
\label{Fig13}
\end{figure}

Figure~\ref{Fig13} compare the retrieved temperature profiles with equilibrium condensation curves to identify regions in which condensates may be thermochemically stable. 
Over the pressures primarily probed by the emission spectra, the dayside profiles remain hotter than most of the displayed condensation curves, making refractory condensates thermochemically unfavourable in the hottest dayside regions. In contrast, near $10^{-2}$~bar, the nightside profiles lie below the condensation temperatures of several refractory species, including Al$_2$O$_3$, CaTiO$_3$, Fe-bearing condensates, and silicates. Condensation-favourable conditions also occur near the limbs, but the retrieved thermal structure is markedly asymmetric: crossings are extensive near $-90^{\circ}$, whereas near $+90^{\circ}$ they are largely confined to higher latitudes. This east--west contrast implies that condensates may preferentially form at the cooler (morning) limb while the warmer (evening) limb remains largely cloud-free. If realised as asymmetric cloud coverage, this dichotomy would produce wavelength-dependent differences between the morning- and evening-terminator transmission spectra by modifying the effective transit radius and the amplitudes of atomic and molecular absorption features. A comparable limb asymmetry has been observed for WASP-76b, where asymmetric Fe absorption was attributed to condensation on the morning limb and scale height differences between limbs \citep{Ehrenreich+etal+2020,Savel+etal+2022,Wardenier+etal+2021}. More recently, rotational-transit observations of WASP-121b revealed longitudinal temperature and compositional gradients through phase-dependent CO and H$_2$O absorption, independently demonstrating that its transmission spectrum varies across the transit \citep{Gapp+etal+2026}. Separately, the lower Bayesian evidence of the \textsc{Cloud} variant relative to \textsc{Base} indicates that the data do not favour the specific globally uniform grey-cloud prescription tested here. This model comparison does not, however, exclude localised or composition-dependent clouds. The combined NIRISS and NIRSpec/G395H dayside analysis of \citet{Saha+Jenkins+2025} reported evidence for CaTiO$_3$ clouds with a high mass fraction. However, the CaTiO$_3$ condensation curve calculated assuming solar composition does not intersect our retrieved dayside temperature structure (see Figure~\ref{Fig13}), suggesting that such clouds may require either enhanced gas-phase CaTiO$_3$ abundances or dynamical transport of nightside condensates to the dayside. This motivates future multidimensional retrievals that jointly analyse phase-resolved G395H spectra with complementary NIRISS observations. Predicting the actual cloud distribution and opacity will ultimately require coupling retrieved three-dimensional temperature structures with cloud microphysics, atmospheric transport, cold trapping, and wavelength-dependent optical properties.

\begin{figure*}[htbp]
\centering
\includegraphics[width=\textwidth]{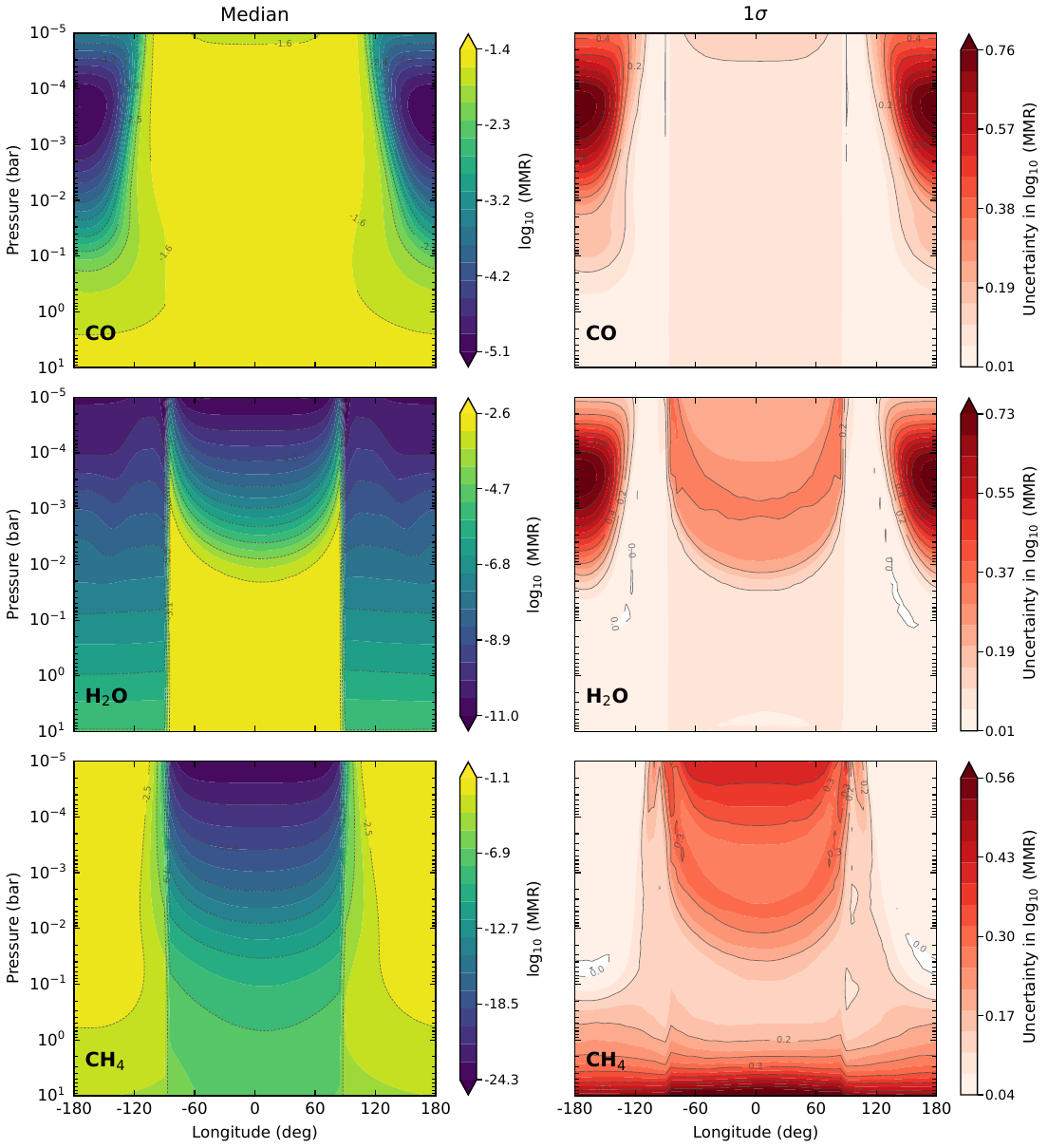}
\caption{
Retrieved equatorial longitude--pressure abundance distributions of key opacity-bearing species from the \textsc{DNChem} model.
The left column shows the posterior median \(\log_{10}{\rm MMR}\), and the right column shows the corresponding \(1\sigma\) uncertainty in \(\log_{10}{\rm MMR}\). 
Rows show CO, H$_2$O, and CH$_4$ from top to bottom. 
The distributions reflect both the independently retrieved dayside and nightside elemental abundances and the local equilibrium response to the temperature--pressure structure. At photospheric pressures, CO and H$_2$O are more abundant over the dayside, whereas CH$_4$ is strongly enhanced on the nightside.
}
\label{Fig14}
\end{figure*}

Figure~\mbox{\ref{Fig14}} shows the derived equatorial longitude--pressure distributions of CO, H$_2$O, and CH$_4$ for the \textsc{DNChem} model. Because \textsc{DNChem} retrieves independent dayside and nightside elemental abundances, these distributions reflect both the hemispheric compositional difference and the response of local chemical equilibrium to the retrieved temperature--pressure structure. Near $10^{-2}$~bar, the substellar CO and H$_2$O abundances are $\log_{10}{\rm MMR}=-1.48$ and $-3.71$, respectively, compared with $-3.80$ and $-7.75$ at the antistellar longitude. CH$_4$ exhibits the opposite behaviour, increasing from $\log_{10}{\rm MMR}=-12.89$ at the substellar point to $-1.53$ on the nightside. H$_2$O is also strongly depleted in the uppermost dayside atmosphere by thermal dissociation. The sharp transitions near the terminators partly reflect the piecewise assignment of independent dayside and nightside elemental abundances and should not be interpreted as resolved chemical discontinuities. The maps therefore require distinct dayside and nightside chemical states within the present equilibrium-chemistry model. Figure~\ref{FigA4} isolates the effects of the local temperature--pressure profile and the adopted hemispheric elemental-abundance. At the antistellar point, the retrieved nightside elemental abundance pattern leads to a substantially higher CH$_4$ abundance than the dayside pattern under the same local temperature–pressure structure, demonstrating that the cooler nightside temperature alone cannot reproduce the inferred methane enhancement. The required effective CH$_4$ enrichment is consistent with disequilibrium chemistry and atmospheric transport proposed for the observed nightside methane \mbox{\citep{Evans-Soma+etal+2025}}, although the present retrieval does not uniquely identify the responsible mechanism. The inferred hemispheric elemental contrasts should therefore be interpreted as a phenomenological representation of different dayside and nightside chemical regimes rather than as a resolved discontinuity in the bulk elemental composition.

\section{Discussion}\label{DISC}

\subsection{Interpretations of WASP-121b's Atmosphere}

The retrieved temperature field is characterised by a hot region approximately centred on the substellar point, a substantially cooler nightside, and a small hotspot offset that increases with pressure. The combination of a large day--night temperature contrast and limited hotspot displacement is broadly consistent with GCMs that include atmospheric drag together with H$_2$ dissociation and recombination \citep{2024MNRAS.528.1016T,Bell+etal+2018}. A general requirement for inhibited zonal heat transport is also supported by the near-zero offset inferred from the Spitzer phase curve \citep{Davenport+etal+2025} and by phase-resolved high-resolution spectroscopy, for which drag is required to reproduce the observed Doppler signatures of CO and H$_2$O \citep{Wardenier+etal+2024,Smith+etal+2024}. Our retrieval adds a distinct vertical constraint to this picture. 
Although the increase in hotspot offset with pressure can partly reflect the longer radiative timescale at depth, the small absolute offsets across the pressures probed by G395H remain below the non-magnetic GCM predictions. This comparison indicates that radiative effects alone are unlikely to account for the strong suppression of the longitudinal displacement and supports the presence of additional atmospheric drag. 
Magnetic drag provides a natural explanation for this additional, pressure-dependent damping because the coupling between the atmospheric flow and the magnetic field depends sensitively on temperature, ionisation fraction, and electrical conductivity.
In the hot, partially ionised atmosphere, the ionisation fraction predicted by the Saha equation increases towards lower pressures at a given temperature, strengthening magnetic coupling and allowing Lorentz forces to suppress zonal motion in the upper atmosphere, whereas weaker coupling at greater pressures may permit more efficient eastward heat transport and a larger hotspot offset \citep{Perna+etal+2010,Rogers+etal+2014}. 
Within the comparison between the $0$, $3$~G, and $10$~G GCMs, the retrieved offsets across the pressures probed by JWST/NIRSpec G395H are more consistent with the $3$~G case. This agreement supports magnetic drag as a plausible interpretation but does not constitute a measurement of the planetary magnetic field. A several-gauss field is broadly compatible with the $\sim2$~G estimate inferred from high-resolution wind measurements \citep{Seidel+etal+2026}. It is also consistent with evolutionary calculations in which Ohmic dissipation reduces the deep convective flux supporting the dynamo and yields surface fields of order $1$--$10$~G \citep{Vigano+etal+2025}, although these predictions depend on the adopted heating prescription and dynamo scaling.  Nevertheless, the present comparison does not uniquely identify the drag mechanism. 

The retrieved temperature near the dayside hotspot reaches approximately $3500$~K, modestly higher than predicted by the GCMs (see Figure~\ref{Fig10}), which may partly reflect the omission or simplified treatment of opacity sources that absorb stellar radiation and heat the low-pressure atmosphere, including atomic metals, TiO/VO, and H$^{-}$, as well as differences in the assumed chemical abundances.

The strong evidence preference for \textsc{DNChem} indicates that a single globally shared elemental composition is inadequate within the present equilibrium chemistry model. It should not be interpreted as evidence for a physical discontinuity in the planet's elemental inventory. Instead, the independently retrieved nightside abundance group provides an effective methane enrichment beyond that produced by the retrieved nightside temperature structure alone. Transport-induced disequilibrium chemistry therefore provides a compelling physical explanation for the enhanced nightside CH$_4$, consistent with the interpretation of \mbox{\citet{Evans-Soma+etal+2025}}. If the dayside remains close to thermochemical equilibrium while the nightside methane abundance is maintained out of equilibrium, the resulting hemispheric contrast would indicate a strong spatial variation in the competition between chemical relaxation and atmospheric transport. Efficient horizontal mixing alone would tend to diminish this contrast, whereas nightside-localised vertical mixing could establish a distinct chemical quench level and maintain a chemical state different from that of the dayside. Although hydrostatic scaling implies that vertical velocities are much smaller than horizontal velocities, \(W/U\sim H/L\ll1\), the chemical importance of vertical motions is not determined solely by their velocity magnitude, but rather by the competition between vertical transport and chemical relaxation timescales. The characteristic vertical transport timescale can be estimated as \(\tau_{\rm vert}\sim H/W\), which may become comparable to or shorter than the chemical relaxation timescale \(\tau_{\rm chem}\). When \(\tau_{\rm chem}\gtrsim\tau_{\rm vert}\), vertical transport can redistribute chemical species before chemical reactions restore local thermochemical equilibrium, leading to chemical quenching and the persistence of disequilibrium abundances. Therefore, even weak vertical motions can exert a significant influence on atmospheric composition when vertical abundance gradients are strong and chemical relaxation is sufficiently slow. Determining whether sufficiently vigorous vertical mixing can be sustained on the nightside will require GCM calculations coupled self-consistently to disequilibrium chemical kinetics. Nevertheless, the retrieved hemispheric abundance groups may also absorb the effects of spatially varying clouds or residual deficiencies in the thermal and radiative-transfer models. The Bayesian preference thus establishes the need for distinct effective dayside and nightside chemical states within the tested model family, while the physical process responsible for maintaining the enhanced nightside methane remains to be determined. 

Ground-based high-resolution spectroscopy offers independent support for the intrinsically 3D atmospheric structure implied by our results. Phase-resolved observations reveal species-dependent absorption and emission trails with distinct Doppler shifts, demonstrating that different molecules trace different spatial distributions and wind projections across the terminator and dayside. For instance, detections of CO, OH, and H$_2$O alongside refractory species shows different velocity offsets \citep{Smith+etal+2024,Wardenier+etal+2024,Pelletier+etal+2025}, consistent with chemical and thermal inhomogeneity arising from temperature-dependent dissociation and circulation effects. This is compatible with our inference that H$_2$O is preferentially suppressed in the hottest regions, while CO remains comparatively uniform (see Figure~\ref{Fig14}). Together, the species-dependent dynamics from high-resolution spectroscopy and the multidimensional structures retrieved here provide mutually consistent constraints on atmospheric dynamics and composition, and offer a framework for future studies that couple circulation, chemical, and cloud processes.

The \textsc{Base} retrieval yields C/H, O/H, and Si/H enrichments of $48.12_{-2.58}^{+1.46}\times$, $27.08_{-1.43}^{+0.80}\times$, and $8.71_{-1.08}^{+1.22}\times$ solar, respectively. The corresponding \textsc{DNChem} dayside values are $6.26_{-0.99}^{+1.33}\times$, $4.18_{-0.62}^{+0.81}\times$, and $7.56_{-1.04}^{+1.23}\times$ solar. \citet{Pelletier+etal+2026}, who likewise defined their volatile and refractory abundance parameters relative to solar, inferred enrichments of approximately $13\times$ and $20\times$ solar when reflected light was excluded, and approximately $17\times$ and $15\times$ solar when reflected light was fitted freely. For comparison with \citet{Evans-Soma+etal+2025}, we normalise our posteriors to their median non-LTE host-star abundances. This gives C/H, O/H, and Si/H enrichments of $38.49_{-2.06}^{+1.17}\times$, $18.52_{-0.98}^{+0.55}\times$, and $5.67_{-0.71}^{+0.80}\times$ stellar for \textsc{Base}, compared with $5.00_{-0.79}^{+1.06}\times$, $2.86_{-0.42}^{+0.56}\times$, and $4.93_{-0.68}^{+0.80}\times$ stellar for the \textsc{DNChem} dayside. \citet{Evans-Soma+etal+2025} reported C/H, O/H, and Si/H enrichments of $23.96_{-4.75}^{+6.13}\times$, $12.19_{-2.24}^{+2.78}\times$, and $9.89_{-2.89}^{+5.99}\times$ stellar. Thus, \textsc{Base} favours higher C/H and O/H but lower Si/H, whereas the \textsc{DNChem} dayside values are lower for all three elements. The retrieved C/O ratios likewise differ among analyses: $\mathrm{C/O}=0.977_{-0.003}^{+0.003}$ for \textsc{Base}, $0.825_{-0.022}^{+0.020}$ for the \textsc{DNChem} dayside, $0.92_{-0.03}^{+0.02}$ in \citet{Evans-Soma+etal+2025}, $0.48_{-0.16}^{+0.14}$ without reflected light and $0.82_{-0.09}^{+0.05}$ with freely fitted reflected light in \citet{Pelletier+etal+2026}, and $0.963\pm0.024$ in \citet{Saha+Jenkins+2025}. A recent panchromatic dayside retrieval combining NIRISS/SOSS, NIRSpec/G395H and MIRI/LRS inferred lower volatile enrichments than \citet{Evans-Soma+etal+2025}, closer to the lower dayside C/H and O/H favoured by \textsc{DNChem}, while still requiring a strongly enriched refractory inventory \citep{Kahle+etal+2026}. This dispersion highlights the sensitivity of retrieved compositions to wavelength coverage and model assumptions. In addition, because only a limited subset of the species considered in the equilibrium-chemistry calculations is included as opacity sources in the radiative-transfer model, omitted absorbers may introduce systematic biases into the retrieved elemental abundances and C/O ratio.

\subsection{Methodological Strengths and Limitations}

\subsubsection{Fidelity of Eclipse Normalisation}

The eclipse normalisation framework presented in this work offers distinct advantages for analysing JWST time-series spectroscopy of exoplanet atmospheres. 

First, by normalising each cleaned integration against the stellar reference template derived from the full secondary eclipses, the method directly generates a phase-resolved residual data cube. This process is computationally efficient, bypassing the light-curve fitting required by traditional pipelines, and is thus ideally suited for characterising longitudinal variations in planetary emission. 

Second, the procedure preserves the native spectral resolution of the instrument. By avoiding aggressive wavelength binning early in the reduction, it enables the extraction of weak planetary signals using high-resolution cross-correlation techniques \citep[e.g.,][]{2023ApJ...955L..19E,2025MNRAS.543.3456E} or retrievals. 

Finally, the reduction stage is model-independent, relying on no prior assumptions regarding the planetary phase curve morphology. This flexibility allows the subsequent retrieval framework to incorporate atmospheric and systematic noise models of arbitrary complexity, enabling a fully data-driven, end-to-end inference atmospheric properties.

However, certain limitations and caveats apply. A primary requirement is high-quality data with sufficient phase coverage, particularly the capture of a clean baseline during full secondary eclipse. The method's effectiveness may be reduced for datasets with sparse temporal sampling or short eclipse durations where the stellar reference is poorly constrained. Furthermore, while the current implementation leverages the dual-eclipse baseline of WASP-121b to detrend linear systematics, it does not implicitly solve for unknown, complex time-dependent noise sources. In principle, physically motivated noise models could be jointly inferred within the retrieval framework to mitigate this. Additionally, potential sources of contamination such as intrinsic stellar variability or non-linear wavelength-dependent instrumental drifts are not explicitly modelled in the current setup. Applying this method to more active stars or observations with longer baselines may require supplementary correction strategies to ensure robust atmospheric characterisation.

\subsubsection{Applicability of the Temperature Parameterisation}
Our temperature parameterisation is designed to reconstruct the atmospheric temperature fields with a minimal number of free parameters while retaining essential physical content. To achieve this, we prioritise capturing dominant large-scale structures and deliberately suppress secondary, small-scale features. This strategy is well suited to the interpretation of exoplanetary spectra, which are spatially unresolved and highly degenerate, encoding primarily global information. Unlike Solar System observations where spatially resolved spectroscopy probes localised atmospheric structures, exoplanet phase curves provide only disk-integrated signals modulated by rotation. Within this observational constraint, our framework aims to assess whether the retrieved large-scale structure aligns with theoretical expectations, and to identify potential deviations that may point to missing physical processes.

\subsubsection{Approximations in Radiative Transfer}
In our current framework, the inherently 3D radiative transfer problem is approximated by summing independent 1D column calculations, and scattering processes are neglected to maintain computational efficiency for retrieval. This approximation may introduce systematic biases in the inferred thermal structure and molecular abundances.

Recent studies \citep[e.g.,][]{Arora+etal+2025} have demonstrated that substantial differences in thermal emission and transmission spectra can arise between 1D and fully 3D radiative transfer treatments, particularly in strongly irradiated atmospheres. 
These deviations likely reflect the limitations of the 1D approximation in our work, which neglects the multidimensional radiative geometry and employs simplified treatments of scattering and geometric projection. This highlights the critical need for developing computationally efficient, fully 3D radiative transfer techniques to robustly interpret high-precision JWST phase curves observations and forthcoming high-resolution spectroscopic measurements from next-generation ground-based facilities.

\section{Conclusions}\label{CONCL}

In this work, we have developed a framework for extracting and interpreting multidimensional atmospheric structures from JWST spectroscopic phase curves and applied it to the ultra-hot Jupiter WASP-121b. Our approach integrates two key components: (1) the eclipse normalisation method, which extracts wavelength-dependent phase-resolved emission spectra directly from the time series without parametric light-curve fitting; and (2) a physically motivated temperature-field parameterisation that links the longitudinal structure to the characteristic radiative, advective, and diffusive timescales. The eclipse-normalised spectra are statistically consistent with those obtained from standard light-curve fitting, while preserving the native instrumental resolution and requiring fewer assumptions at the extraction stage. The analytical temperature parameterisation reproduces the large-scale morphology of representative GCM temperature fields and, when coupled to disk integration of independently calculated 1D radiative-transfer columns, provides a computationally efficient forward model for multidimensional retrieval.

By applying this framework to the JWST/NIRSpec G395H phase-curve observations of WASP-121b, we have derived a coherent picture of its global thermal structure. 
The preferred model with independently retrieved dayside and nightside elemental abundances (\textsc{DNChem} model) exhibits a pronounced day--night temperature contrast, a dayside thermal inversion, and thermal inversions at both limbs, with the eastern and western limb temperatures differing by several hundred Kelvin. The model comparison further indicates that models allowing nightside thermal inversions are favoured over those strictly prohibiting them, suggesting that at least part of the nightside atmosphere hosts a thermal inversion. Dynamically, the vertical profiles of the dimensionless parameters indicate a regime transition: the upper atmosphere is radiation-dominated, while deeper layers show an increasing influence of advection and diffusion. 
The hotspot offset increases across the pressures probed by G395H, from $4.0^{\circ}$ near $10^{-3}$~bar to $8.7^{\circ}$ near $0.1$~bar for \textsc{DNChem}, compared with $1.2^{\circ}$ to $3.7^{\circ}$ for the cloud-free model with globally shared elemental abundances (\textsc{Base}). 
The confined dayside hot region and small, pressure-dependent hotspot offsets favour a magnetically damped GCM over a corresponding non-magnetic calculation, indicating that additional atmospheric drag is required beyond radiative effects alone. Magnetic drag provides a natural interpretation, although alternatives such as Rayleigh drag cannot be excluded. Measurements of the pressure-dependent wind structure may help break the degeneracy among drag mechanisms.

The retrieved chemical distributions exhibit longitude-dependent variations that follow the large-scale temperature structure. 
\textsc{DNChem} is strongly preferred over \textsc{Base}, indicating that a globally uniform elemental abundance prescription is insufficient to describe the observed atmosphere. The retrieved hemispheric elemental abundance differences are best interpreted as distinct effective chemical states rather than a discontinuity in the bulk elemental inventory, with transport-induced disequilibrium chemistry providing a plausible explanation for the enhanced nightside CH$_4$. The retrieved thermal structures further suggest spatially varying cloud formation conditions, with asymmetric limb temperatures potentially leading to different condensation regimes and distinct transmission spectra \citep{Gapp+etal+2026}. However, stronger constraints on cloud properties require broader wavelength coverage (e.g., NIRISS). Together, these results highlight the importance of multidimensional chemistry, clouds, and circulation in shaping the observed atmosphere of WASP-121b.

Overall, the retrieved chemical properties are broadly consistent with previous JWST analyses of WASP-121b. The elemental enrichments inferred by \textsc{Base} are comparable in magnitude to those reported by \citet{Evans-Soma+etal+2025} and \citet{Pelletier+etal+2026}, although the inferred composition depends sensitively on the adopted chemical parameterisation. In addition, the C/O ratio changes from $0.977_{-0.003}^{+0.003}$ in \textsc{Base} to $0.825_{-0.022}^{+0.020}$ for the \textsc{DNChem} dayside. For comparison, \citet{Evans-Soma+etal+2025} reported $\mathrm{C/O}=0.92_{-0.03}^{+0.02}$, whereas \citet{Pelletier+etal+2026} obtained $0.48_{-0.16}^{+0.14}$ when reflected light was excluded and $0.82_{-0.09}^{+0.05}$ when its contribution was fitted freely. The \textsc{DNChem} nightside solution instead characterises an effective disequilibrium chemical state and should not be interpreted as a direct constraint on the bulk elemental inventory. Quantitative inferences regarding atmospheric metallicity and planetary formation history therefore remain conditional on the adopted chemical model.

The eclipse normalisation approach is intended as a complementary quick-look extraction rather than a complete replacement for traditional light-curve fitting. A fully rigorous treatment would incorporate explicit modelling of instrumental systematics within the retrieval framework, enabling a more comprehensive propagation of uncertainties and potential degeneracies. Collectively, the eclipse normalisation technique and the physically motivated parameterisation offer an efficient, physically interpretable pathway for constraining the 3D nature of exoplanet atmospheres. The methodology is readily applicable to future phase-curve observations with \textit{JWST} and \textit{Ariel}, paving the way for unified multidimensional retrievals capable of disentangling the complex interplay of circulation, chemistry and energy transport in strongly irradiated worlds. Extending the framework to fully 3D radiative transfer, disequilibrium chemical kinetics, spatially resolved cloud microphysics, and self-consistent magnetohydrodynamic GCMs will be essential for testing the physical origin of the nightside methane enrichment, cloud spatial distribution and pressure-dependent atmospheric damping.

\begin{acknowledgments}

We thank the anonymous referee for constructive comments that improved the manuscript. We thank Johannes U. Lange for advice and guidance on the use of \textsc{NAUTILUS} and the assessment of retrieval convergence. We thank James S. Fecanin and Robert Collingwood Frazier for their support with the GCM calculations, and Lile Wang for helpful discussions on hydrodynamics.
G.C. acknowledges the support by the National Natural Science Foundation of China (NSFC; grant Nos. 42578016, 12122308, 42075122), Youth Innovation Promotion Association CAS (2021315), and the Minor Planet Foundation of the Purple Mountain Observatory. 
All of the data presented in this article were obtained from the Mikulski Archive for Space Telescopes (MAST) at the Space Telescope Science Institute. The specific observations analysed can be accessed via \dataset[doi:10.17909]{https://doi.org/10.17909/6qnn-6j23}.
The associated data products and source code are publicly available on Zenodo at \url{https://doi.org/10.5281/zenodo.21534046}.

\end{acknowledgments}

\bibliography{ref.bib}{}
\bibliographystyle{aasjournal}




\appendix

\section{Performance comparison between \texttt{MultiNest} and \texttt{NAUTILUS}}\label{app_A}

\begin{figure}
\centering
\includegraphics{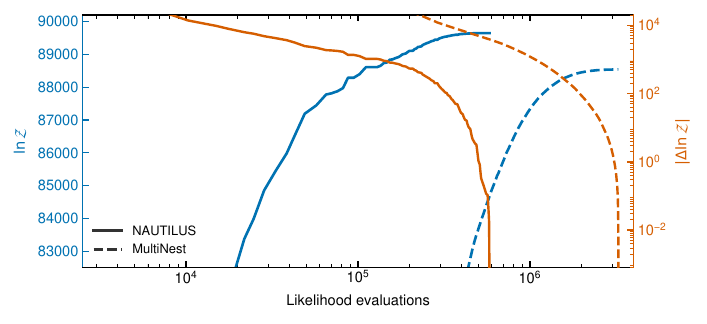}
\caption{
Performance comparison between \texttt{MultiNest} and \texttt{NAUTILUS} for the multidimensional retrieval. The left axis shows the $\ln \mathcal{Z}$, while the right axis shows the $|\Delta \ln \mathcal{Z}|$ as a function of the number of likelihood evaluations. \texttt{NAUTILUS} achieves a given evidence precision with substantially fewer likelihood evaluations than \texttt{MultiNest}, demonstrating its higher sampling efficiency for the present high-dimensional retrieval problem.
}
\label{FigA1}
\end{figure}

To assess the computational performance of nested-sampling algorithms in our high-dimensional atmospheric retrieval framework, we compared a representative  \texttt{MultiNest} retrieval with an equivalent \texttt{NAUTILUS} retrieval for the \textsc{Base} model using the evidence-based stopping diagnostic shown in Fig.~\ref{FigA1}. The \texttt{MultiNest} run required $3.26\times10^{6}$ likelihood evaluations to satisfy the stopping criterion, whereas the \texttt{NAUTILUS} run required only $5.81\times10^{5}$ evaluations. Thus, for this retrieval problem, \texttt{NAUTILUS} reduced the number of likelihood evaluations by a factor of $5.6$, corresponding to an $82\%$ reduction in forward-model calls.

Figure~\ref{FigA1} shows the evolution of the Bayesian evidence ($\ln \mathcal{Z}$) and its difference relative to the final evidence estimate ($|\Delta \ln \mathcal{Z}|$) as a function of the number of likelihood evaluations. Owing to its deep-learning-assisted proposal distribution design \citep{Lange+2023}, \texttt{NAUTILUS} achieves a given evidence precision substantially more efficiently than \texttt{MultiNest}. Given the computational expense of the multidimensional retrieval framework and the extensive model comparisons performed in this work, all retrievals presented in the main text were carried out using \texttt{NAUTILUS}.

\section{Additional figures}

\begin{figure*}[htb!]
\centering
\includegraphics[width=\textwidth]{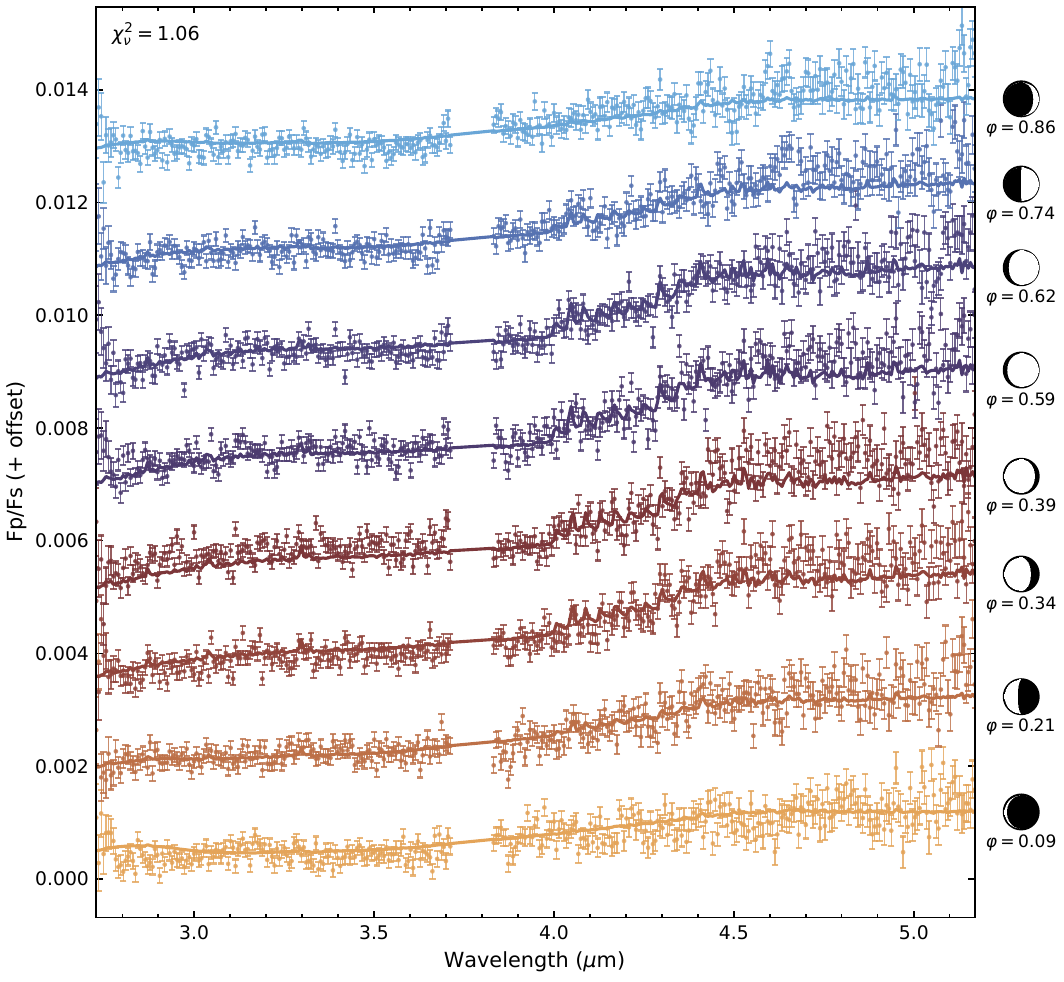}
\caption{
Phase-resolved emission spectra of WASP-121b at representative orbital phases for the \textsc{DNChem} retrieval. 
The points with error bars show the observed planet-to-star flux ratios after excluding the transit and secondary-eclipse intervals. 
Solid curves show the corresponding best-fitting model spectra. 
For clarity, the spectra are vertically offset, with the orbital phases indicated on the right. 
The model captures the systematic increase in thermal emission from nightside-dominated phases to dayside-dominated phases across the NIRSpec/G395H bandpass.
}
\label{FigA2}
\end{figure*}

\begin{figure*}[htbp]
\centering
\includegraphics[width=\textwidth]{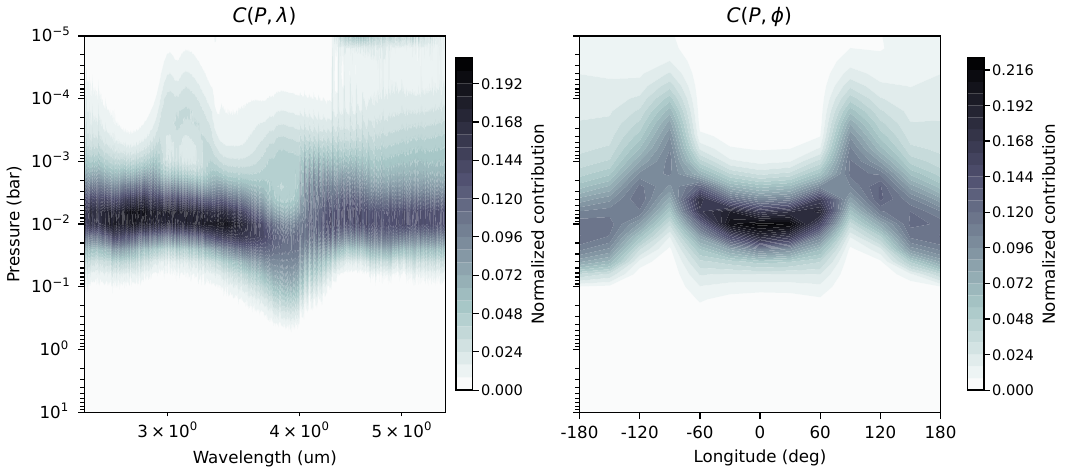}
\caption{
Contribution functions for the \textsc{DNChem} model.
The pressure--wavelength contribution map shows the atmospheric pressures probed across the NIRSpec/G395H bandpass, while the pressure--longitude contribution map shows how the sensitivity varies with longitude after integrating over wavelength.
}
\label{FigA3}
\end{figure*}

\begin{figure*}[htbp]\centering\includegraphics[width=\textwidth]{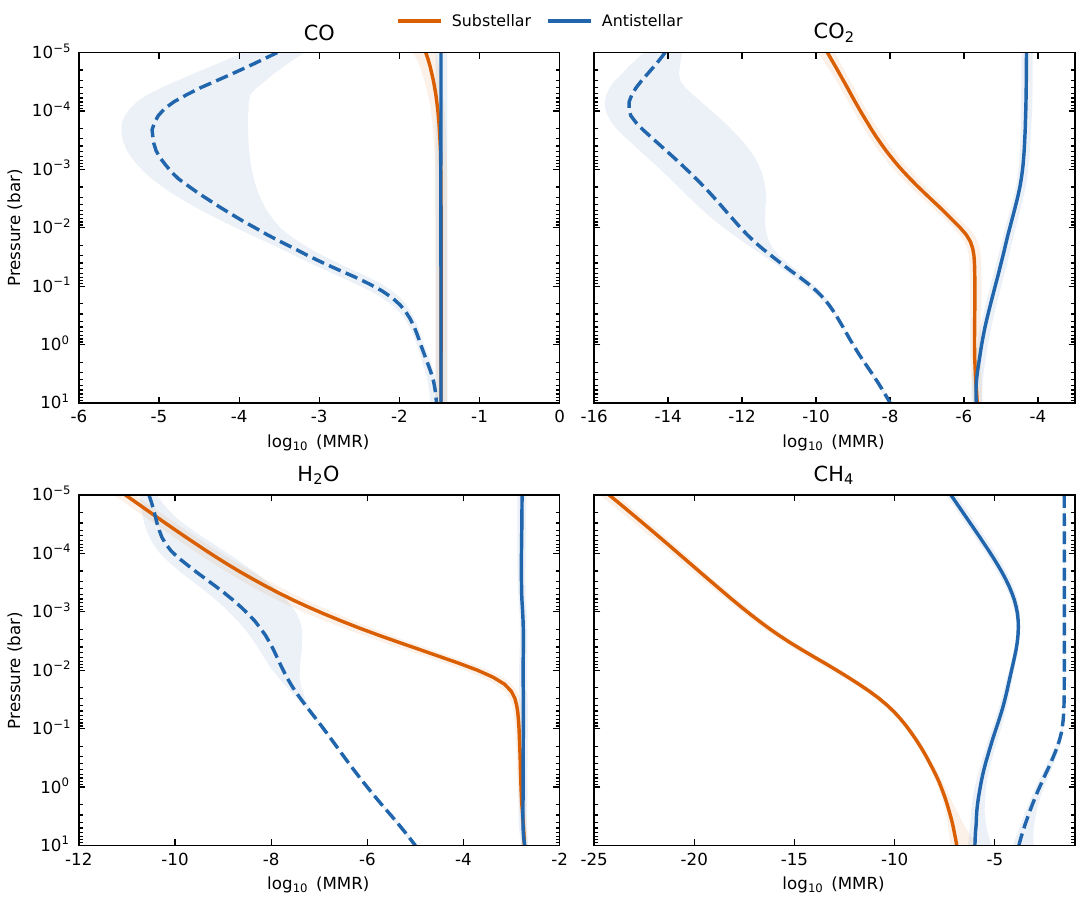}
\caption{
Mass-mixing-ratio profiles of CO, CO$_2$, H$_2$O, and CH$_4$ implied by the \textsc{DNChem} posterior. Orange and blue curves are evaluated at the substellar and antistellar points, respectively. Solid curves adopt the dayside elemental abundances, whereas dashed curves adopt the nightside elemental abundances. The curves show the posterior medians, and the shaded regions denote the $1\sigma$ uncertainties. At the antistellar point, the nightside abundance group produces substantially more CH$_4$ than the dayside group at the same local temperature--pressure structure, demonstrating that the inferred nightside methane enhancement cannot be reproduced by the cooler temperature alone. This effective enrichment is consistent with nightside disequilibrium chemistry and atmospheric transport.}
\label{FigA4}
\end{figure*}

\begin{figure*}[htbp]
\centering
\includegraphics[width=\textwidth]{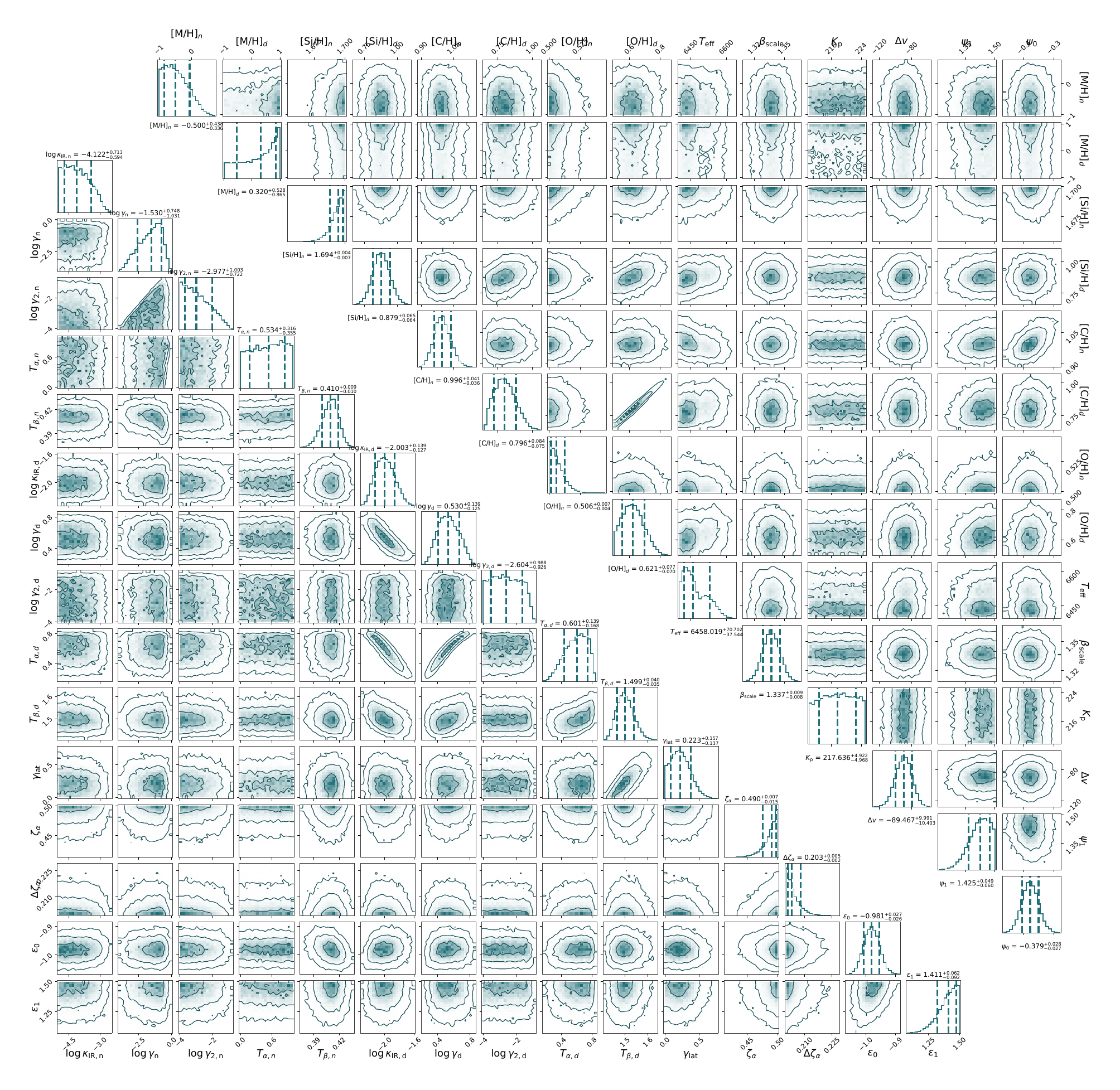}
\caption{
Posterior distributions of the retrieved parameters for the \textsc{DNChem} model, which assigns independent elemental abundances to the dayside and nightside while retaining local chemical equilibrium within each hemisphere. 
}
\label{FigA5}
\end{figure*}

\begin{figure*}[htbp]
\centering
\includegraphics[width=\textwidth]{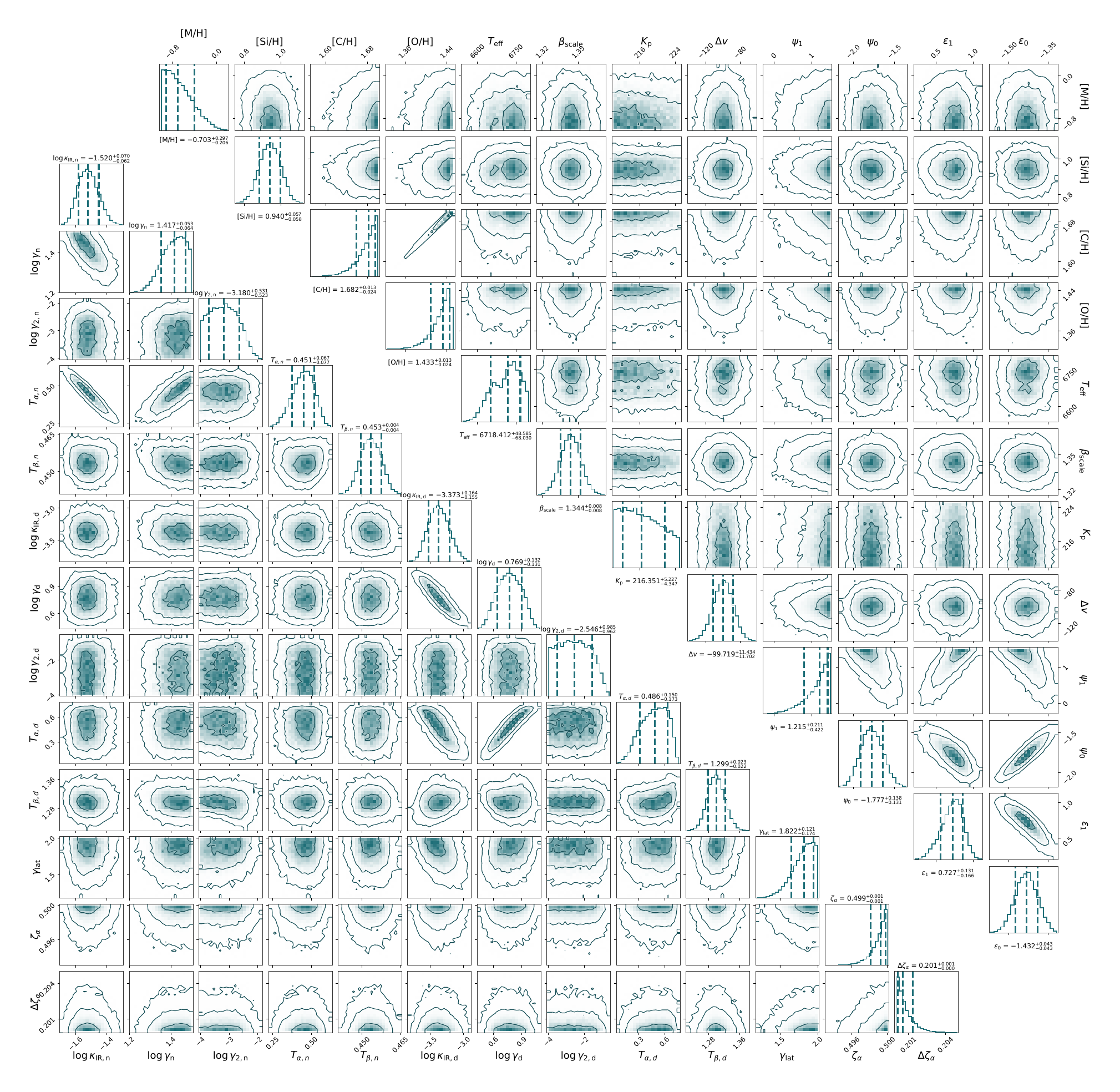}
\caption{
Posterior distributions of the retrieved parameters for the \textsc{Base} model, which assumes a cloud-free atmosphere with globally shared elemental abundances. 
}
\label{FigA6}
\end{figure*}

\end{document}